\documentclass[10pt]{article}

\usepackage[utf8]{inputenc} 
\usepackage[T1]{fontenc}    
\usepackage{url}            
\usepackage{booktabs}       
\usepackage{nicefrac}       
\usepackage{microtype}      

\usepackage{amsmath,amssymb,amsfonts,textcomp,graphicx,nicefrac,mathtools,mathrsfs, dsfont, amsthm, enumitem} 
\usepackage{dsfont}

\usepackage{bbm}
\usepackage[utf8]{inputenc}

\usepackage[title]{appendix}
\usepackage{soul}

\usepackage{mathtools}
\usepackage{footmisc}
\usepackage{tikz}

\usepackage{wrapfig}

\usepackage{xcolor}
\usepackage{float,caption}
\usepackage[colorlinks=true, linkcolor=red, citecolor=blue]{hyperref}

\makeatletter
\def\BState{\State\hskip-\ALG@thistlm}
\makeatother

\newtheoremstyle{dotless}{}{}{\itshape}{}{\bfseries}{}{ }{}
\theoremstyle{dotless}

\newtheorem*{myprop*}{Proposition}
\theoremstyle{plain}
\newtheorem{myth}{Theorem}

\newtheorem{mylem}[myth]{Lemma}

\newtheoremstyle{named}{}{}{\itshape}{}{\bfseries}{.}{.5em}{#1 #3}
\theoremstyle{named}
\newtheorem*{namthm*}{Theorem}

\makeatletter
\newenvironment{shortlem}{%
    \refstepcounter{myth}
    \textbf{Lemma~\themyth.}%
    \space\em}{%
    \par}
\makeatother

\newtheorem{innercustomgeneric}{\customgenericname}
\providecommand{\customgenericname}{}
\newcommand{\newcustomtheorem}[2]{%
  \newenvironment{#1}[1]
  {%
   \renewcommand\customgenericname{#2}%
   \renewcommand\theinnercustomgeneric{##1}%
   \innercustomgeneric
  }
  {\endinnercustomgeneric}
}

\newcustomtheorem{customthm}{Theorem}
\newcustomtheorem{customlemma}{Lemma}

\newcommand{\ip}[1]{\left\langle #1 \right\rangle}
\newcommand{\SNR}{\textrm{SNR}}

\usepackage[ruled]{algorithm2e}
\usepackage[noend]{algpseudocode}

\makeatletter
\def\BState{\State\hskip-\ALG@thistlm}
\makeatother

\usepackage{subcaption}

\usepackage[
backend=bibtex,
maxcitenames = 1,
style=alphabetic,
sorting=anyt,
maxbibnames=99
]{biblatex}
\usepackage{geometry}

\newcommand{\tpose}{\mathsf{T}}
\renewcommand{\exp}[1]{\operatorname{exp}\left(#1\right)}
\newcommand{\qedee}{\Box}
\usepackage{import}

\title{\Large{Testing Changes in Communities for the Stochastic Block Model}}
\author{ 
\large{Aditya Gangrade} \\ \normalsize{Boston University} \\ \normalsize{\texttt{gangrade@bu.edu}} 
\and
\large{Praveen Venkatesh}\\ \normalsize{Carnegie Mellon University} \\ \normalsize{\texttt{vpraveen@cmu.edu}} 
\and
\large{Bobak Nazer} \\ \normalsize{Boston Univeristy} \\ \normalsize{\texttt{bobak@bu.edu} }
\and
\large{Venkatesh Saligrama} \\ \normalsize{Boston University} \\\normalsize{ \texttt{srv@bu.edu} }} \date{\vspace{-12pt}}

\begin{document}

\maketitle

\begin{abstract}

We propose and analyze the  problems of \textit{community goodness-of-fit and two-sample testing} for stochastic block models (SBM), where changes arise due to modification in community memberships of nodes. Motivated by practical applications, we consider the challenging sparse regime, where expected node degrees are constant, and the inter-community mean degree ($b$) scales proportionally to intra-community mean degree ($a$). Prior work has sharply characterized partial or full community recovery in terms of a  ``signal-to-noise ratio'' ($\mathrm{SNR}$) based on $a$ and $b$. For both problems, we propose computationally-efficient tests that can succeed far beyond the regime where recovery of community membership is even possible. Overall, for large changes, $s \gg \sqrt{n}$, we need only $\mathrm{SNR}= O(1)$ whereas a na\"ive test based on community recovery with $O(s)$ errors requires $\mathrm{SNR}= \Theta(\log n)$. Conversely, in the small change regime, $s \ll \sqrt{n}$, via an information-theoretic lower bound, we show that, surprisingly, no algorithm can do better than the na\"ive algorithm that first estimates the community up to $O(s)$ errors and then detects changes. 
We validate these phenomena numerically on SBMs and on real-world datasets as well as Markov Random Fields where we only observe node data rather than the existence of links.
\end{abstract}



While community detection and recovery for the stochastic block model (SBM) \cite{abbe2017community} and, more generally, inference of community structures underlying large-scale network data~\cite{girvan2002community,newman2006modularity,fortunato2010community} has received significant interest across the machine learning, statistics and information theory literatures, there has been limited work on the important problem of testing changes in community structures. 
%
The general problem of testing changes in networks naturally arises in a number of applications such as discovering statistically significant topological changes in gene regulatory networks~\cite{Zhang09} or differences in brain networks between healthy and diseased individuals~\cite{Bassett}. Building upon this perspective, we propose testing of differences in the underlying community structure of a network, which can encompass scenarios such as detecting structural changes over time in social networks \cite{adamic2005political,fortunato2010community}, determining whether a set of genes belong to different communities in disease and normal states \cite{jiang2004cluster}, and deciding whether there are changes in functional modules, which represent communities, in protein-protein networks \cite{chenBo_PPI}. 

Testing structural changes in networks is statistically challenging due to the fact that we may have relatively few independent samples to evaluate combinatorially-many potential changes. In this paper, we propose methods for goodness-of-fit (GoF) testing and two-sample testing (TST) for detecting changes in community memberships under the SBM. The SBM naturally captures the community structures commonly observed in large-scale networks, and serves as a baseline model for more complex networks. Specifically, there are $n$ nodes partitioned into two equal-sized communities, and the network is observed as a random $n\times n$ adjacency matrix, representing the instantaneous pairwise interactions among individuals in the population. Both intra- and inter-community interactions are allowed. Members within the same community interact with uniform probability $a/n$, while members belonging to different communities with a smaller probability $b/n$. We restrict attention to the commonly-considered and practically-relevant setting of $a/b = \Theta(1)$.

For our testing problems, we assume that the network samples are aligned on $n \gg 1$ vertices, and that the latent communities are either the same, or they differ in at least some $s \ll n$ nodes. We pose the GoF problem as: \emph{Decide whether or not the observed random incidence matrix is an instantiation of a given community structure.} For the TST problem, we ask: \emph{Given two random incidence matrices, decide whether or not their latent community structure is identical.}

\paragraph{Sparse vs. Dense Graphs.} We focus on scenarios where the observed random incidence matrices are sparse with average node degree bounded by a constant independent of the network size. Within this context we develop minimax optimal methods for GoF and TST in this context. We are motivated by both practical and theoretical concerns. Practically, as observed in~\cite{chung2010graph}, realistic graphs such as social networks are sparse (friendships do not grow with network size); in temporal settings, at any given time, only a small subset of interactions are observed; and in other cases ascertaining the presence or absence of each edge in the network being observed is an expensive process, and it makes sense to understand the fundamental limits for when testing is even possible. 

From a theoretical standpoint, the sparse setting is challenging due to signal-to-noise ratio (SNR) constraints that do not arise in the dense case. Recovery of the latent community with up to $s$ errors is possible iff $\Lambda \gtrsim \log(n/s)$ \cite{chin2015stochastic, zhang2016minimax, fei2019exponential}, where $\Lambda$ is a SNR parameter that, in the setting $a/b = \Theta(1),$ scales linearly with the mean degree. In particular, for $\Lambda$ of constant order, recovery with sublinear distortion fails. The question of \emph{whether testing is possible when recovery fails} is mathematically intriguing. Further, this is the \emph{only} theoretically interesting setting. Indeed, if testing for $s$ changes requires a graph dense enough to allow recovery with $\sim s$ errors, then one might as well recover these communities and compare them.  

\paragraph{Our Contributions.} We show that optimal tests exhibit a surprising two-phase behavior: \begin{enumerate}[wide]
\item For $s \gg \sqrt{n}$, or `large changes,' we propose computationally-efficient schema for GoF and TST that succeed with $\Lambda = O(1)$ - far below the SNR threshold for recovery. For GoF, this requirement is even weaker - we only need $\Lambda \gtrsim n/s^2,$ which vanishes with $n$ since $s \gg \sqrt{n}$. Further, we match these bounds up to constants with information-theoretic lower bounds.
\item In contrast, we show via an information-theoretic lower bound that for $s \ll \sqrt{n}$, or `small changes,' both testing problems require $\Lambda = \Omega(\log(n))$ for reliable testing. This means that the na\"{i}ve strategy of recovering communities and comparing them is tight up to constants in this regime.\end{enumerate}

We complement the above theoretical study by three experiments: the first implements the above tests on synthetic SBMs, and the second on the political blogs dataset - a popular real world dataset for community detection \cite{adamic2005political}. Both of these experiments show excellent agreement with the theoretical predictions. The third experiment casts a wider net, and instead studies the related problem of testing the underlying community structure of a Gaussian Markov Random Field that has precision matrix $I + \gamma G$ for $G$ drawn from an SBM. This experiment explores the more realistic setting where instead of receiving a graph, we obtain observations at each node of a hidden graph, and wish to reason about the underlying structure. Remarkably, a simple adaptation of our procedure for SBMs shows excellent performance for this problem. This indicates that our observations are not restricted to raw SBMs, but may signal a more general phenomenon that merits exploration.

\paragraph{Related Work.} For work on recovery communities we refer to the survey \cite{abbe2017community}. However, we explicitly point out the papers \cite{chin2015stochastic, zhang2016minimax, fei2019exponential}, which provide various schemes and necessary conditions that show that the partial recovery problem with distortion $s$ can be solved with vanishing error probability if and only if $\Lambda \gtrsim \log(n/s)$. We further point out the lower bounds of \cite{Mossel2015lowerbound,deshpande2015asymptotic}, which assert that if $\Lambda < 2,$ then asymptotically, the best possible distortion for partial recovery (or weak recovery, as it is referred to in this constant SNR regime) is $n/2 - o(n)$. Note that reporting a uniformly random community achieves distortion of $s = n/2 - O(\sqrt{n}).$

Ours is the first work to study GoF and TST where both hypothesized models are SBMs. Nevertheless, both GoF and TST in the context of network data as well as SBMs have been studied. Below we highlight the key differences in modeling assumptions and the ensuing technical implications, which renders much of the prior work inapplicable to our setting. 

With regards to GoF, \cite{arias2014community,verzelen2015community} study the problem of detecting if a graph is an unstructured Erd\H{o}s-R\'{e}nyi (ER) graph, or if it has a planted dense subgraph, providing detailed characterizations of the feasiblity regions and statistical phase transitions in this setting. While this work is aligned with ours in the techniques used, the modeled setting and problem there are different (ER vs. planted dense subgraph), and TST is not explored. Particularly, the dense subgraph model and the SBM are qualitatively different, and conclusions from one cannot be transferred to the other directly. 

A number of papers, including \cite{lei2016goodness, bickel2016hypothesis, banks2016information, gao2017testing} study various techniques and regimes of determining if a graph is a SBM or an unstructured ER graph, and if the former, the number of communities in the model. Of these, \cite{gao2017testing} approach the problem by counting small motifs in the graphs, \cite{banks2016information} propose a simple scan and \cite{lei2016goodness, bickel2016hypothesis} propose testing of the number of communities on the basis of the top singular values of the graph. 

\cite{Tang2017TST} study TST of the model parameters in random dot product graphs, and propose the distance between aligned spectral embeddings of the two graphs as a statistic to do so. They use this to test equality against various transformations of the underlying models, and in particular for SBMs, test if the connectivity probabilities $(a/n,b/n)$ are identical or not for two graphs with latent communities that are randomly drawn. \cite{li2018two} adapt these tests by considering the same distance, but weighted by the corresponding singular values of one of the graphs, and use this to study two-sample testing of equality of the latent communities in the graphs - as in this paper.

In contrast to the low-rank structure assumptions in the above work, \cite{ghoshdastidar2017two, pmlr-v65-ghoshdastidar17a, ghoshdastidar2018practical} study two-sample testing of inhomogeneous ER graphs (i.e., ER graphs where each edge may have a distinct probability of existing). Within this setting, they provide a number of statistics based both on estimates of the Frobenius and operator norms of the differences of the expected graph adjacency matrices, as well as those based on  motifs such as triangles, and explore the limits of these tests. 

A fundamental drawback of these approaches, in our context, is their reliance on singular values, spectral norms and Frobenius norms. Singular embeddings are particularly sensitive to noise, and stable embeddings require significant edge density (particularly when a sublinear number of alterations to the communities are to be tested). Indeed, in this context, we note that, in contrast to our low SNR, sparse setting, \cite{li2018two} require both a degree of $n^{1/2 - \epsilon}$ and an SNR of $\log(n)$  corresponding to a high SNR, high edge-density regime, where full community recovery is possible.

Similarly, Frobenius and Spectral norms based tests of \cite{ghoshdastidar2018practical,ghoshdastidar2017two} are not stable enough to test a sublinear number of changes in a low SNR regime. Functionally, this can be seen by the fact that the square-Frobenius norm of the difference of two graphs is equal to the number of edges that appear in one graph but not the other, and for sparse graphs, \emph{most} edges appear in only one of the two graphs. Similarly, arguments about spectral norms rely on concentration of the same for ER graphs, but the best known concentration radius \cite{le2017concentration} is far too large to allow testing of small differences in sparse graphs. Indeed, for any of the statistics of \cite{ghoshdastidar2018practical} to have power in our setting, the results of the paper require that the expected degree diverges with $n$, and that $\Lambda \gtrsim n/s,$ which is exponentially above the SNR required to recover communities up to distortion $s/2$.

\section{Definitions}\label{sec:defi}

\paragraph{The Stochastic Block Model.} A vector $x \in \{\pm 1\}^n$ is said to be a \emph{balanced community vector} (or partition) if $\sum x_i = 0$. The \emph{stochastic block model} is defined as a random, simple, undirected graph $G$ on $n$ nodes such that all edges are drawn mutually independently given $x$, and \[ P( \{i,j\} \in G|x) = \frac{a+b}{2n} + \frac{a-b}{2n} x_i x_j. \] Note that we treat $x$ as a deterministic but unknown quantity, and thus, $P(\cdot|x)$ is a slight abuse of notation. The parameters $(a,b)$ may vary with $n$, and we focus on the setting  $a,b = O(\log n),$ with emphasis on $O(1)$\footnote{{While our main interest is in the constant degree regime, we also show that testing for small changes is impossible in this setting (e.g Thm \ref{thm:gof_bal}), and instead logarithmic degrees are needed. Thus, to present our results completely, we must allow $a,b$ to vary at least in the range $[\Omega(1), O(\log(n))],$ or, more succinctly $O(\log n)$. Large scales are not of interest since exact recovery is possible at the logarithmic scale.}}, and $a/b = \Theta(1)$. For technical convenience, we require that $a+b < n/4$. 

\noindent The \emph{signal-to-noise ratio} (SNR) of an SBM is the quantity $\displaystyle \Lambda := \frac{(a-b)^2}{a+b},$ which characterises the recovery problem, as described in earlier discussions.

\noindent Note that the partitions $x$ and $-x$ induce the same distribution. Accordingly, the \emph{distortion} between partitions $x$ and $y$ is $ d(x,y) := \min( d_{\mathrm{H}} (x,y), d_{\mathrm{H}} (x,-y) ),$ where $d_{\mathrm{H}}$ is the Hamming distance. \\

\noindent\textbf{Minimax Testing Problems.} We formally define two minimax hypothesis testing problems.

\paragraph{Goodness-of-Fit.} We are given a balanced partition $x_0$ and a parameter $s.$ We receive a graph $G \sim P(G |x)$, where $x$ is an unknown balanced partition that is either exactly equal to $x_0$ or differs in at least $s$ places. Our goal is to solve the hypothesis test: 
\begin{align*}
        &H_0: d(x,x_0) = 0 \textrm{~~~~~~~vs.~~~~~~~} H_1: d(x,x_0) \ge s 
    \end{align*} We measure the minimax risk of this problem by 
    \begin{equation}
        R_{\mathrm{GoF}}(n,s,a,b) := \adjustlimits \inf_{\phi} \sup_{x_0} \Big\{ P(\textrm{FA}) + \, \sup_{{x}} \, P(\textrm{MD}(x))\Big\} \label{gofminimaxrisk}
    \end{equation}
    where $\phi(G)$ outputs either $0$ or $1$, $P(\textrm{FA}):=P(\phi(G)=1\mid x_0)$, $P(\textrm{MD}(x)):= P(\phi(G) = 0\mid x)$, and the second supremum is over all $x$ such that $d(x,x_0) \geq s$. \\
    
\paragraph{Two-Sample Testing.} We are given a parameter $s$ and two independent graphs $G \sim P(G| x), H \sim P(H| y),$ where $x$ and $y$ are unknown balanced communities satisfying $d(x,y) \in \{0\} \cup [s:n/2].$  The goal is to solve the following (composite null) testing problem:
    \begin{align*}
        H_0: d(x,y) = 0 \textrm{~~~~~~~vs.~~~~~~~} H_1: d(x,y) \ge s,
    \end{align*}
     with the measure of risk \begin{equation} 
R_{\mathrm{TST}}(n,s,a,b) := \adjustlimits \inf_{\phi} \sup_{x,y} P\big( \phi(G,H) \neq \mathbf{1}\{ x = y\} \,|\, x,y \big), 
\end{equation} where $\phi(G,H)$ outputs either $0$ or $1$ and the supremum is over balanced $x,y$ such that $d(x,y) \in \{0\} \cup [s:n/2].$\\

\noindent \textbf{On Reliability:} As we vary $n$ and $(s,a,b)$ with $n$ as some functions $(s_n, a_n, b_n)$, the above define a sequence of hypothesis tests. We say that the GoF problem can be solved \emph{reliably} for such a sequence if $R_{\mathrm{GoF}}(n,s_n,a_n, b_n) \to 0$ as $n\nearrow \infty,$ and similarly for TST. Below, we will target $O(1/n)$ bounds. For conciseness, we will suppress the dependence of risks on $(n,s,a,b),$ writing just $R_{\mathrm{GoF}}/R_{\mathrm{TST}}$.\\

{\noindent \textbf{On balance:} The strict balance assumption above can be relaxed to only requiring that both communities are of size linear in $n$, at the cost of weakening some of the constants left implicit in the theorem statements. While the majority of the analysis in the paper will assume exact balance, we briefly discuss unbalanced but linearly sized communities whilst detailing the proofs. Note that since the communities are no longer balanced, the differences between $x$ and $y$ can be `one-sided' i.e., more nodes can move from, say, $+$ to $-$, than in the other direction. We do not require any control on these other than the total number of changes is at least $s$.} \\

\noindent \textbf{On constants:} We use $C$ and $c$, and their modifications, as unspecified constants that may change from line to line. While these can be explicitly bounded, we do not expect them to be tight.  

\section{Community Goodness-of-Fit}\label{sec:gof}

We begin by stating our main results regarding the \textit{community goodness-of-fit problem}.

\begin{myth}\label{thm:gof_bal} Community goodness-of-fit testing is possible with risk $R_{\mathrm{GoF}}\le \delta$ if $s\Lambda \ge C \log(2/\delta)$ and $ \displaystyle \Lambda \ge C \frac{n}{s^2} \log(2/\delta) $ for some constant $C > 0$. 

Conversely, in order to attain $R_{\mathrm{GoF}} \le \delta \le 0.25,$ we must have that $s\Lambda \ge C' \log(1/\delta)$ and $\displaystyle 
 \Lambda \ge C' \log \left( 1 + \frac{n}{s^2} \right)  $ for some constant $C' > 0$.
\end{myth}

\noindent These bounds reveal the following behavior in terms of large and small changes:\\

\noindent \textbf{Large changes.} If $s \geq n^{1/2 + c}$ for some $c > 0$, then, since $n/s^2 \le 1$ and $\log(1+x) \ge x/2$ for $x \le 1,$ the second converse bound behaves as $\Lambda \ge C n/s^2.$ This matches the sufficient condition up to a constant.

\noindent \textbf{Small changes.} On the other hand, if $s \leq n^{1/2 - c}$ for some $c > 0$,  since $n/s^2 \sim n^{2 c}$, the second converse bound instead behaves as $\Lambda \gtrsim \log n$. In this regime, community recovery up to $s/2$ errors requires $\Lambda \geq C \log{2n/s} = \tilde{C} \log{n}$. Thus, estimating $x$ from $G$ and comparing it to $x_0$ is optimal up to constants.

\noindent \textbf{Transition.} The above indicate a phase transition in the GoF testing problem at $\sigma:= \log_{n}(s) = 1/2.$ Consider the thermodynamic limit of $n \nearrow \infty.$ For $\sigma < 1/2,$ the problem is `hard' in that the SNR $\Lambda$ is required to diverge to $\infty$, while for $\sigma > 1/2,$ the SNR can tend to zero.

\begin{proof}[{Proof Sketch for the Achievability}]\let\qed\relax Let us begin with an intuitive development of the test. Since we start with a partition $x_0$ in hand to test, it is natural to look at the edges across and within the cut defined by $x_0.$ We thus define the number of edges \emph{across} and \emph{within} this cut:
 	\begin{equation}\label{eqn:NaNw} \begin{aligned}
 		N_a^{x_0}(G) &:= |\{ (i,j) \in G: x_{0,i} \neq x_{0,j} \}|  = \frac{1}{4}x_0^\tpose (D(G) - G) x_0 \\ 
 		N_w^{x_0}(G) &:= |\{ (i,j) \in G: x_{0,i} = x_{0,j} \}| = \frac{1}{4} x_0^\tpose  (D(G) + G)x_0 \end{aligned}
 	\end{equation} 
	where the latter expressions treat $G$ as an adjacency matrix and $D(G) = \mathrm{diag}( \mathrm{degree}(i) )_{i \in [1:n]}$.
	\footnote{Note that $D(G) - G$ is the Laplacian of the graph.}
In the null case, these are respectively $\mathrm{Bin}(n^2/4,b/n)$ and $\mathrm{Bin}(2\binom{n}{2},a/n)$ random variables, while in the alternate case some $s/2 \cdot (n-s)/2$ of each behave like edges of the opposite polarity (i.e. as $b/n$ instead of $a/n$ and vice versa), leading to a excess/deficit of edges of this type. Note that while the `average signal strength', i.e., the amount by which edges are over- or underrepresented is the same in both cases ($\sim s|a-b|$), the group with the larger null parameter suffers greater fluctuations. Thus, we base our test only on edges of smaller bias. This reduces the SNR by at most a factor of $4$.
 
We now define the test. $C_1$ below is the constant implicit in Lemma \ref{lem:bal_gof_guts} in Appendix \ref{appx:gof_bal_ach}.
\begin{itemize}
	\item If $a > b,$ we use the test \( \displaystyle N_a^{x_0} (G) \overset{H_1}{\underset{H_0}{\gtrless}} \phantom{-} \frac{bn}{4} \phantom{-} + C_1 \max\left( \sqrt{nb \log(1/\delta)}, \log(1/\delta) \right). \)
	\item If $b > a,$ we use the test  \(\displaystyle N_w^{x_0} (G) \overset{H_1}{\underset{H_0}{\gtrless}} \frac{an}{4} - \frac{a}{2} + C_1 \max\left( \sqrt{na \log(1/\delta)}, \log(1/\delta) \right). \) 
\end{itemize}

The risks of these tests can be controlled by separating the null and alternate ranges using Bernstein's inequality. Indeed, the threshold above is just the the expectation plus the concentration radius of the statistic under the null distribution. Let us briefly develop the statistic's behaviour in the alternate - considering only the case $a>b$, we find that under the alternate, $\binom{n-s}{2} + \binom{s}{2}$ of the edges in $N_a^{x_0}$ continue to behave like $\mathrm{Bern}(b/n)$ bits, while the remaining $s(n-s)/2$ edges behave as $\mathrm{Bern}(a/n)$ bit. Thus, the expectation of $N_a^{x_0}$ is increased by an amount greater than $ s(n-s)\frac{a-b}{2n} \ge s(a-b)/4. $ Next, Bernstein's inequality controls the fluctuations at scale $\sqrt{\max(nb, s(a-b) ) \log(2/\delta)}.$ The conclusion is straightforward to draw from here, and the proof is carried out in Appendix \ref{appx:gof_bal_ach}. \end{proof}

\begin{proof}[{Proof Sketch for the Converse.}]\let\qed\relax The proof is relegated to Appendix \ref{appx:gof_conv}, and we discuss the strategy here. The converse proof follows Le Cam's method, which lower bounds the minimax risk by the Bayes risk for conveniently chosen priors - which can be expressed using the TV distance.

To show $\Lambda \gtrsim \log(1 + {n}/{s^2})$, we pick the null $x_0$ to be any balanced community, and choose the uniform prior on communities that are exactly $s$-far from $x_0$ (in fact, we only use a subset of these in order to facilitate easier computations). This is an obvious choice for this setting - we are interested in balanced communities that are at least $s$ far, and choosing a large number of them allows for a greater `confusion' in the testing problem due to a richer alternate hypothesis. The bound follows by invoking inequalities between $\mathrm{TV}$ and $\chi^2$ divergences and a lengthy calculation due to the combinatorial objects involved. 

To show $s\Lambda \gtrsim -\log(\delta),$ we again pick the null to be any balanced community, and pick the alternate to be an $s$-far singleton. It then proceeds to control $d_{\mathrm{TV}}$ by the Hellinger divergence. \end{proof}

\section{Two-Sample Testing} \label{sec:tst}

We again begin with the main results on \textit{community two-sample testing problem}.

\begin{myth}\label{thm:tst}

Assume, for some $\gamma > 0$, $s \geq n^{\frac{1}{2} + \gamma}$. There exist constants $C, C'$ such that if $C' \le a,b \le (n/2)^{1/3}$, then two-sample testing of $s$ changes with $R_{\mathrm{TST}} \le 4/n$ is possible if the SNR satisfies $\Lambda \ge C$.

\noindent Conversely, for $n \ge 200,$ there exist constants $c, c'$ such that if $ s < (\frac{1}{2} - c')n,$ then two-sample testing of $s$ changes cannot be carried out with $R_{\mathrm{TST}} \le 1/4$ unless $\Lambda \ge c$. 
\end{myth}

\noindent\textbf{Large Changes.} The above theorem makes an achievability claim for the setting of large changes. Notice that in this regime the stated upper and lower bounds match up to constants. Specifically, if $ n^{\frac{1}{2} + \gamma} < s < (\frac{1}{2} - c')n$, two-sample testing can be solved iff $\Lambda \gtrsim 1$. Further, the condition $a,b \gtrsim 1$ is also tight, as it follows from $a/b = \Theta(1)$, and the necessary condition $\Lambda \gtrsim 1,$ since $\Lambda \le a+b$.

This leaves the condition $\max(a,b) \le (n/2)^{1/3}$, which we suspect is an artifact of the proof technique and conjecture that, even for our proposed test, it can be removed. In any case, observe that this condition is irrelevant in the setting $a,b = O(\log n)$ considered in this paper. Further, if $a/b$ is bounded away from $1$, then TST is directly possible when $a,b = \Omega(\log n)$ by recovering the communities and comparing them, demonstrating that this condition is not present in general. \vspace{3pt}\\ 
\noindent \textbf{Small Changes.} We claim that for small changes - $s < n^{\frac{1}{2} - \gamma}$ for some $\gamma > 0$ - the na\"{i}ve scheme of recovering the communities and comparing them is minimax. To see this, note that that GoF testing is reducible to TST - given a TST scheme of a known risk, one may construct a GoF tester of that risk by feeding the TST algorithm the observed graph and a graph drawn from $P(\cdot|x_0)$. Thus, the lower bounds of Theorem \ref{thm:gof_bal} apply to TST, and for $a/b = \Theta(1),$ we find that it is necessary that $s\Lambda = \omega(1)$ and that $\Lambda \gtrsim \log(1 + n/s^2)$ to attain vanishing $R_{\mathrm{TST}}$. For small $s$, the latter lower bound is $\Omega(\log n)$, the claim follows since recovery with up to $s$ errors is possible if $\Lambda \gtrsim \log n.$ \vspace{3pt}\\
\noindent \textbf{Efficiency.} Finally, we point out that the above bounds can be attained with computationally efficient tests. Further, for large changes, the test can be made agnostic to knowledge of $(a,b).$ Instead, it only requires one to be able to estimate $n(a+b)$ to within an additive error of $\widetilde{O}(\sqrt{n(a+b)}),$ which can be done by simply counting the number of edges in the graphs.

\begin{proof}[{Proof Sketch of the Achievability.}]\let\qed\relax We describe the proposed test, and sketch its risk analysis below, completing the same in Appendix \ref{appx:tst_ach}. Recall the definition of $N_w^z, N_a^z$ from (\ref{eqn:NaNw}) in \S\ref{sec:gof}, and let \begin{equation} 
    T^{\hat{x}}(G) := N_w^{\hat{x}}(G) - N_a^{\hat{x}}(G). \end{equation} 
    We show that the routine `TwoSampleTester' below attains a risk smaller than $4/n$. In words, the test computes a partition $\hat{x}$ for the graph $G$ by using about half the edges in the graph. This is represented in the `PartialRecovery' step below, for which any such method may be used - concretely, that of \cite{chin2015stochastic}. Next, we compute the statistic $T^{\hat{x}}$ above for both the remaining part of the first graph, and for the second graph. Notice that unlike the GoF statistic, which was only $N_a,$ $T^{\hat{x}}$ takes the difference of $N_a$ and $N_w.$ This is necessary because the partition $\hat{x}$ derived from partial recovery cannot be very well correlated with the true partition $x$. This means the reduced fluctuations from only considering one part does not apply, and we instead use the whole cut.
   
        \begin{algorithm}[th]                
        \SetCustomAlgoRuledWidth{0.43\textwidth}  
        \caption{TwoSampleTester($G,H, \delta$) }\label{alg:tst}
        \begin{algorithmic}[1]
            \State $G_1 \gets$ subsampling of edges of $G$ at rate $\nicefrac{1}{2}$ uniformly at random.
            \State $\widetilde{G} \gets G - G_1.$
            \State $\widehat{x} \gets \textrm{PartialRecovery}(G_1)$.
            \State Compute $T^{\widehat{x}} (\widetilde{G}), T^{\widehat{x}}(H).$
            \State $T \gets | 2 T^{\widehat{x}} (\widetilde{G}) - T^{\widehat{x}}(H)|.$
            \State Return \( \displaystyle T \underset{H_0}{\overset{H_1}{\gtrless}}  \sqrt{C n (a+b) \log(6n) }.\) 
        \end{algorithmic}
        \end{algorithm}

    Since the edges within communities, and across communities in the graph are (separately) exchangable, the errors made in $\hat{x}$ distribute uniformly over the two communities\footnote{For a proof: since $x, -x$ induce the same law, and since the communities are balanced, for every realization of $G$ such that $\hat{x}$ makes $e_+, e_-$ errors in the community $+,-$ respectively, there is a realization of equal probability where it makes $e_-, e_+$ errors. Further, within community exchangability implies that errors distribute uniformly. }. This allows us to explicitly control the behaviour of $T$ as defined in the test \emph{provided $\hat{x}$ is non-trivially correlatd with $x$} - i.e., given that it makes $< (\nicefrac{1}{2} - c) n$ errors for some $c > 0.$ The condition $\Lambda \gtrsim 1$ in the theorem arises from this.
  
    A complication in this strategy is that the remaining graph $\widetilde{G}$ in the scheme is not independent of the recovered community $\hat{x}.$ This is handled in the analysis by introducing an independent copy of $G$, called $G'$, and arguing that $T^{\hat{x}}( {\widetilde{G} } ) \approx \nicefrac{1}{2} T^{\hat{x}}(G').$ This step is the origin of the nuisance condition $\max(a,b) \lesssim n^{1/3}$ in the theorem. 

    {Lastly, we point out that while the above exploits the exact balance by using the description of the error distribution it enables, one can derive the same results (but with weakened constants) even without this assumption, so long as both communities are at of size linear in $n$. In this case, one cannot rely on the errors distributing uniformly over the nodes, but the within-community uniformity of errors, which follows due to within community exchangability, can be exploited in a similar way. We describe this extension in Appendix \ref{appx:tst_imbalance}.} \end{proof}
    
\begin{proof}[{Proof Sketch of the Converse.}]\let\qed\relax The necessary condition is shown via Le Cam's method, but with the twist that the null model is chosen to be a two-step procedure - one that draws a balanced community uniformly at random, and then generates a graph according to it, while the alternate models are drawn uniformly from the balanced communities that are at least $s$-far from the chosen null. This allows a comparison to the unstructured Erd\H{o}s-R\'{e}nyi graph on $n$ vertices with mean degree $(a+b)/2$. Bounds can then be drawn in from the study of the so-called distinguishability problem \cite{banks2016information}, and we invoke results from \cite{wu2018statistical} to show that total variation distance between the null and alternate distributions is small when $\Lambda$  is a small enough constant, allowing us to conclude using Neyman-Pearson. See Appendix \ref{appx:tst_lower} for a detailed argument. \end{proof}

\section{Experiments} \label{sec:exp}
We perform three different sets of numerical experiments. We first run our tests on  SBMs with $1000$ nodes. Next, we demonstrate that our tests perform similarly for a real dataset, specifically the Political Blogs dataset \cite{adamic2005political}. Finally, we examine SBM-supported Gaussian Markov Random Fields (GMRFs) as an example of a ``node observation'' model, where the SBM-generated edges form the precision matrix for the Gaussian vector consisting of the random variables assigned to each node. In particular, we need to determine if the underlying community of the graph has changed without explicitly observing (or recovering) the edges of the graph. For the sake of brevity, precise details of the experiments are moved to Appendix~\ref{appx:exp}.

\subsection{SBM Experiments} \label{sec:exp_sbm}
\begin{figure}[H]
    \centering
    \includegraphics[width = 0.45\linewidth]{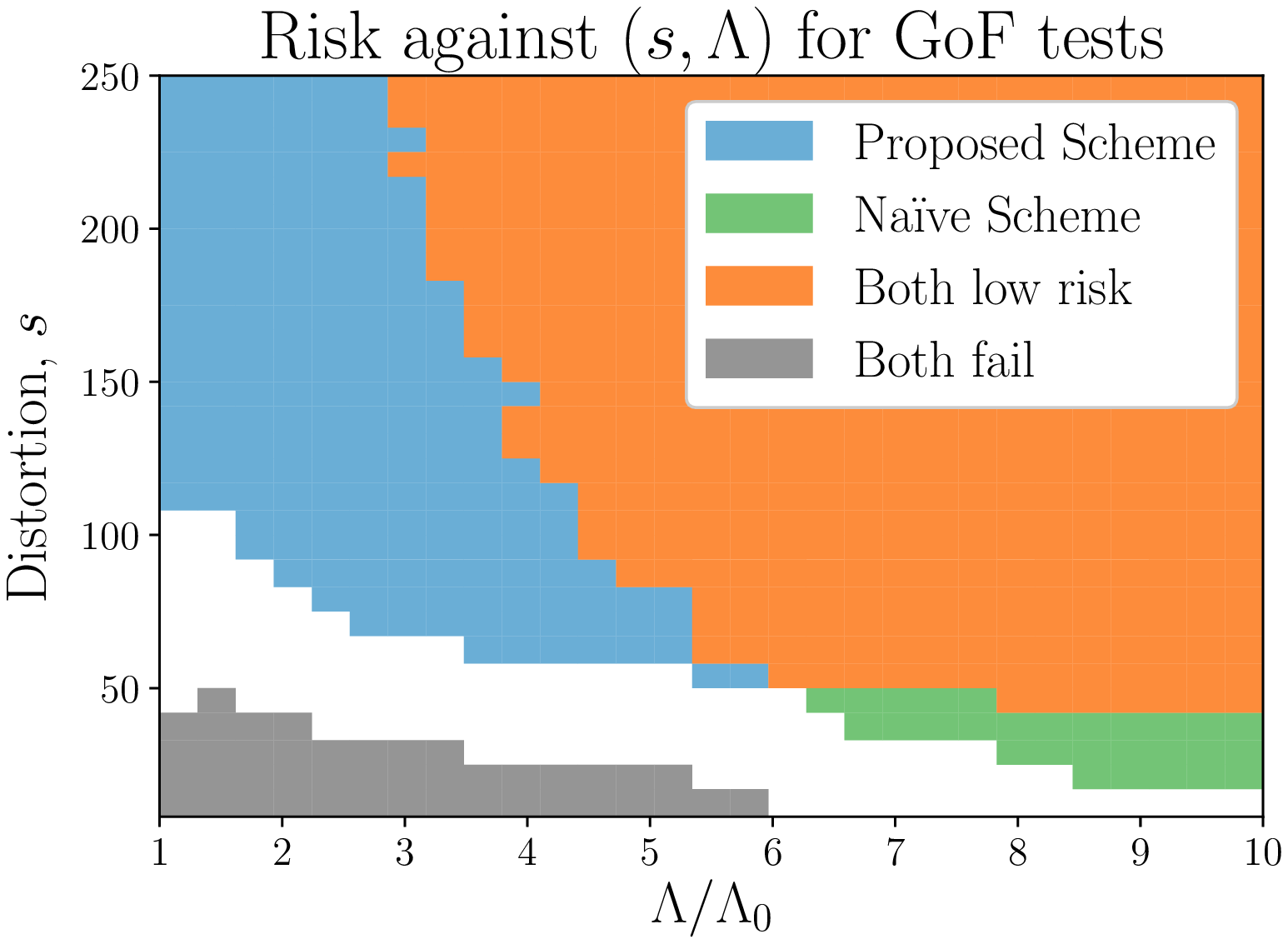}
    ~
    \includegraphics[width = 0.45\linewidth]{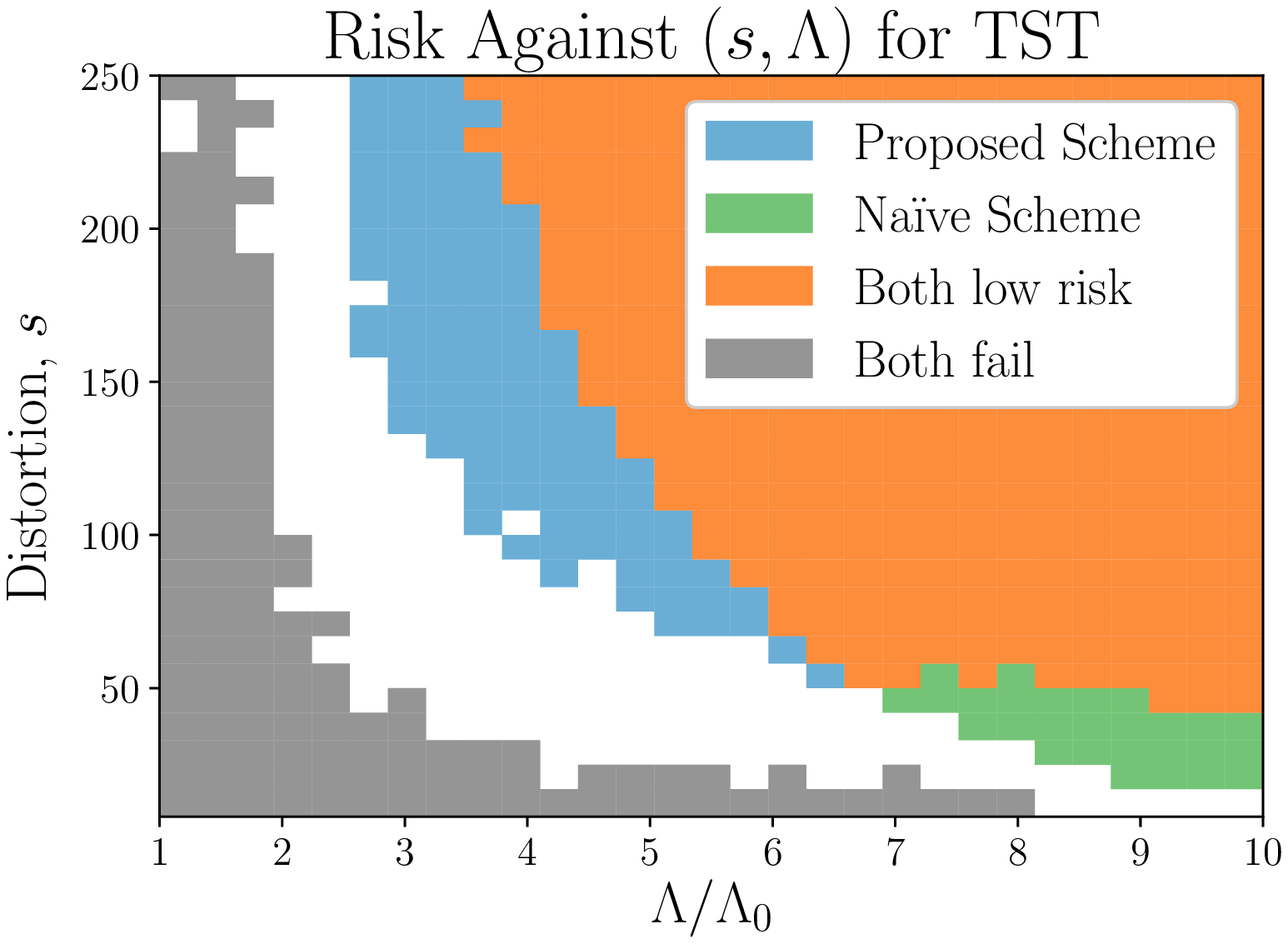}
    \caption{Risks of the proposed tests from Sections \ref{sec:gof} and \ref{sec:tst} for GoF and TST respectively, and the performance of the na\"{i}ve scheme, on synthetic SBMs with $n = 1000, a/b = 3.$ Both schemes attain high risk $(> 1-\delta)$ in the grey region, intermediate risk in the white, and the colours indicate which of the schema attain low risk $(< \delta)$, where $\delta = 0.01$ for GoF and $\delta = 0.1$ for TST.}
    \label{fig:sbm}
\end{figure}

We perform experiments implementing our GoF and TST strategies as well as the na\"{i}ve scheme of reconstructing communities and comparing. Recovery is performed by regularised spectral clustering, for which a detailed description is given in Appendix \ref{appx:exp_sbm}. The graphs are drawn on $n =1000$ nodes for a range of $(s,\Lambda)$ pairs and the high and low risk regimes are plotted in Figure~\ref{fig:sbm}. First, note that for `large changes,' $s \ge \sqrt{n\log(10)}\approx 50$, our GoF and TST tests can succeed for lower SNR values. In contrast, for `small changes,' $s < \sqrt{n} \approx 30,$ the na\"ive test is more powerful in the high SNR regime. Additionally, both tests fail for TST unless the SNR is larger than a constant, as predicted by our lower bound in Theorem~\ref{thm:tst}. 
\subsection[Political Blogs Dataset]{Political Blogs Dataset \cite{adamic2005political}} \label{sec:exp_poli}
The political blogs dataset~\cite{adamic2005political} is canonical in the study of community detection, and consists of $n = 1222$ nodes. Here, we vary the effective SNR by randomly subsampling the edges of the graphs at rate $\rho$. See Appendix \ref{appx:exp_poliblog} for further details. In this dataset, the ground truth partition $x_{\mathrm{True}}$ is available, which in turn yields accurate estimates of the connectivity probabilities $(a,b)$. For this graph $a/b \approx 10$. Further, spectral clustering alone incurs $\approx 50$ errors in this graph, which is larger than $\sqrt{1222} \approx 35$. As a consequence, the behaviour in the `small changes' regime where the test relies on recovery - is not well illustrated in the following. \vspace{3pt}\\
\begin{figure}[ht]
    \centering
    \includegraphics[width = 0.45\linewidth]{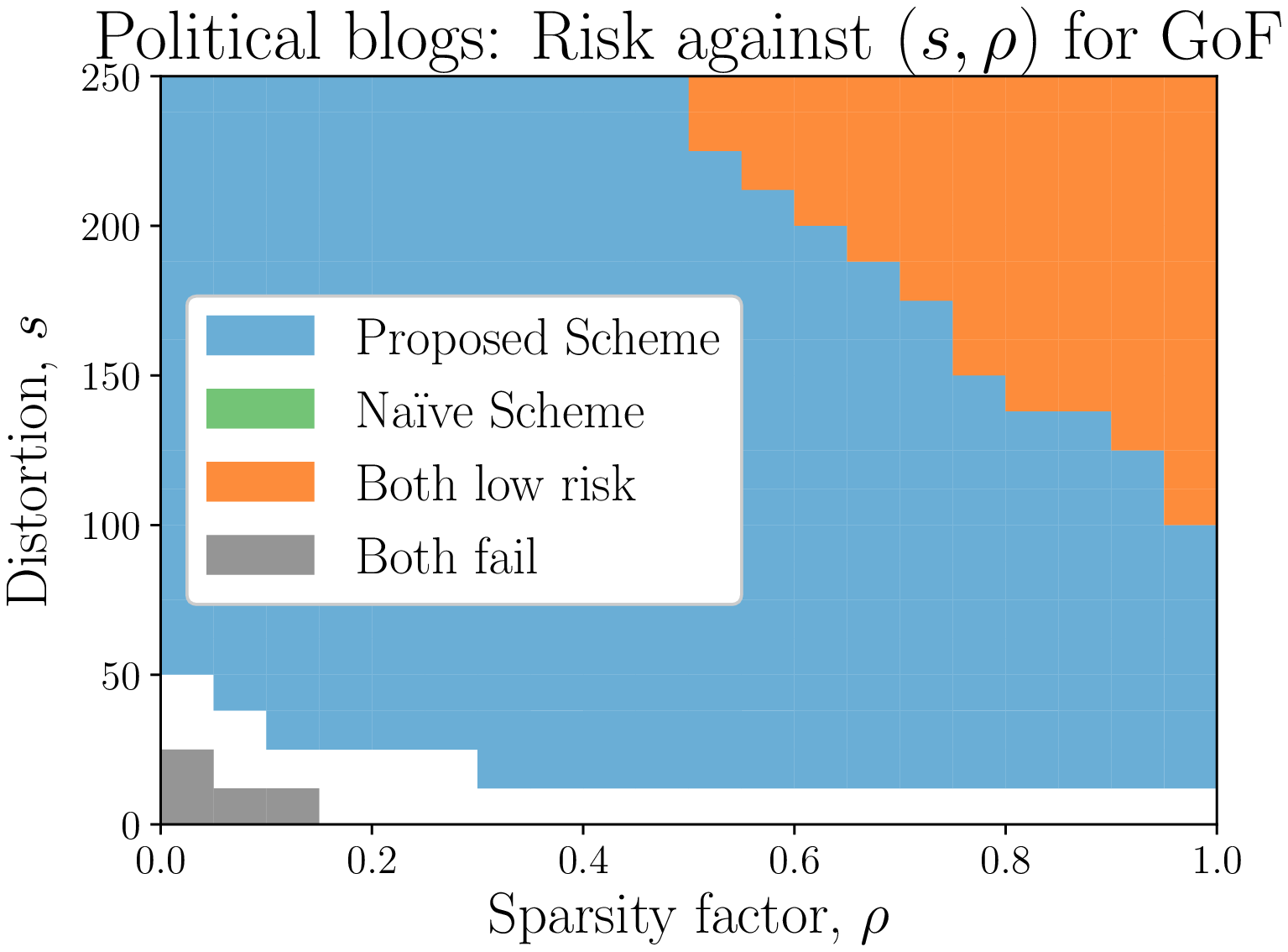}
    ~
    \includegraphics[width = 0.45\linewidth]{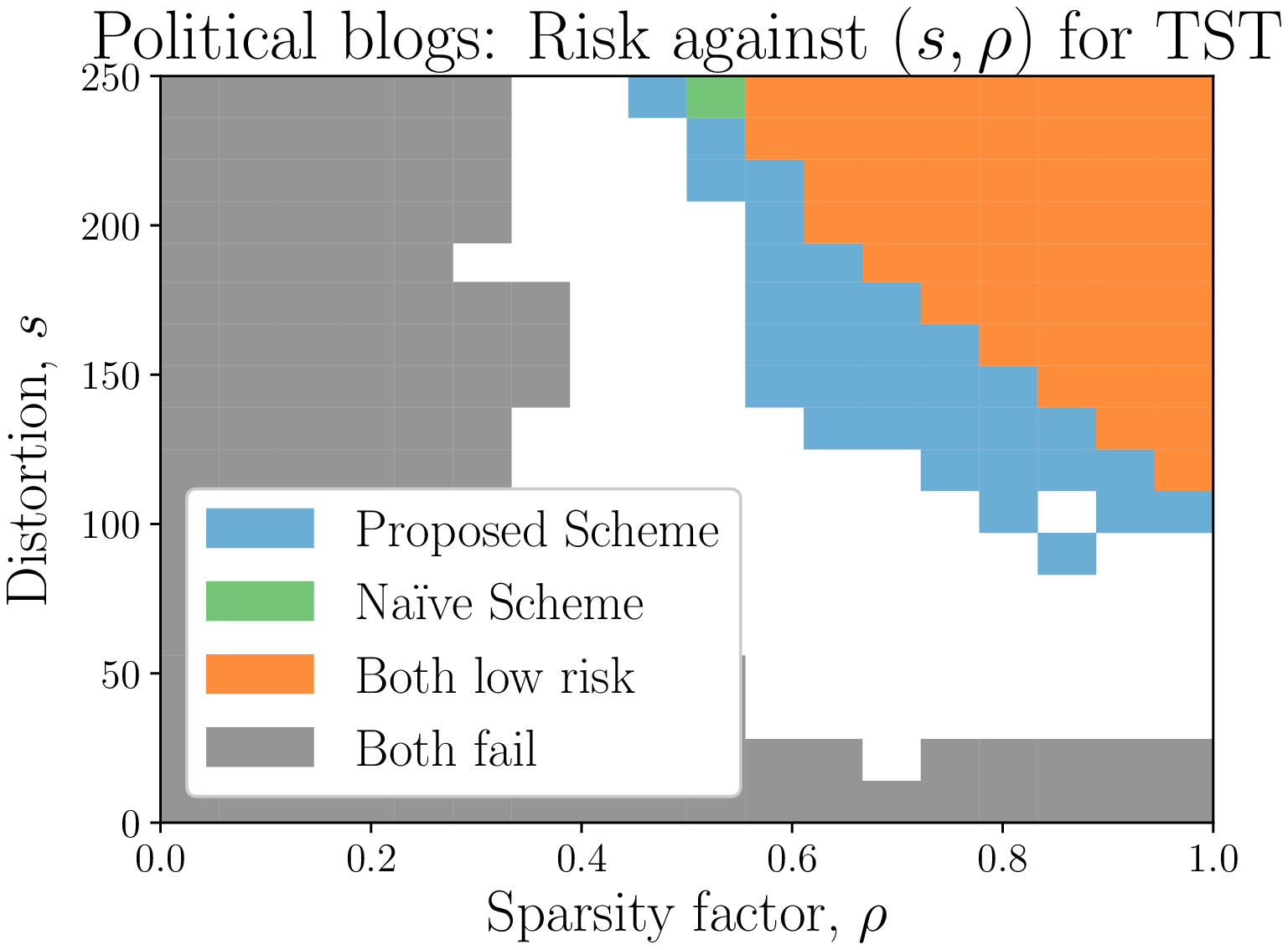}
    \caption{Risks of the tests applied to the Political Blogs graphs - colour scheme is retained from Fig. \ref{fig:sbm}. The X-axis plots the sparsification factor, which serves as a proxy for SNR. Features similar to Fig. \ref{fig:sbm} can be seen. The GoF plot improves since $a/b$ is bigger, while the TST plot suffers since the political blogs graph is not completely described as a 2-community SBM \cite{lei2016goodness}.}
    \label{fig:poli-blogs}
\end{figure}

\noindent \textbf{Goodness-of-Fit.} We determine the size of the test by running the GoF procedures against $x_\mathrm{True}$. To determine power, we construct a partition $y$ by relabelling a random set of nodes of size $s$, and running the GoF procedures against $y$ \emph{with the same graph}.\vspace{3pt}\\
\noindent \textbf{Two-Sample Testing.} We compare the political blogs graph $G$ against two other graphs drawn from SBMs. Size is detemined by drawing $G'$ according to an SBM of community $x_{\mathrm{True}}$ and running the TST procedure, and power is determined by drawing a $y$ as above, generating $H$ according to an SBM of community $y$, and running the TST procedure. Note that this experiment is thus semi-synthetic.
\subsection{Gaussian Markov Random Fields (GMRFs)}\vspace{-8pt}

Frequently instead of simply receiving a graph, one receives i.i.d.~samples from a graph-structured distribution, and it is of interest to be able to cluster nodes with respect to the latent graph. For example, in large-scale calcium imaging, it is possible to simultaneously record the activity pattern of thousands of neurons, but not their underlying synaptic connectivity~\cite{pnevmatikakis2016simultaneous}. Here, we explore the behavior of our tests for GMRFs where the underlying graph structure is randomly drawn from an SBM and and we only observe the nodes. 
\label{sec:exp_gmrf}


A heuristic reason for why our methods might succeed in such a situation arises from the local tree-like property of sparse random graphs  (see, e.g.~\cite{dembo2010gibbs}). For graphs with mean degree $d \ll n,$ typical nodes do not lie in cycles shorter than $\sim \frac{\log n}{2\log d}$. In MRFs, this tree-like property induces correlation decay: the correlation between two nodes decays geometrically up to graph-distance $\sim\frac{\log n }{2\log d}$. Thus, the covariance matrix closely approximates $\sigma_1 G + \sum_{i = 2}^k (\sigma_1 G)^i + \sigma_0 \mathbf{1}\mathbf{1}^\tpose$ for some $\sigma_0 \ll \sigma_1$, small $k$, and $G$, the adjacency matrix of the graph. Since the local structure of the graph is so expressed, both clustering and testing applied directly to the covariance matrix should be viable.

We report experimentation on the GMRF (see, e.g.~\cite[Ch.~3]{MAL-001}), which comprises random vectors $\zeta \sim \mathcal{N}(0, \Theta^{-\!1}),$ where the non-zero entries of the precision matrix $\Theta$ encode the conditional dependence structure of $\zeta.$ Following standard parametrisations \cite{wang2010information}, we set $\Theta = I + \gamma G,$ where $G \sim P(G|x)$ is an adjacency matrix from an SBM with latent parameter $x$, and $\gamma$ is a scalar. Below, we fix the SBM parameters $a,b$ and the level $\gamma$, and explore risks against $s$ and sample size $t$.

\begin{figure}[H]
        \includegraphics[width=0.45\textwidth]{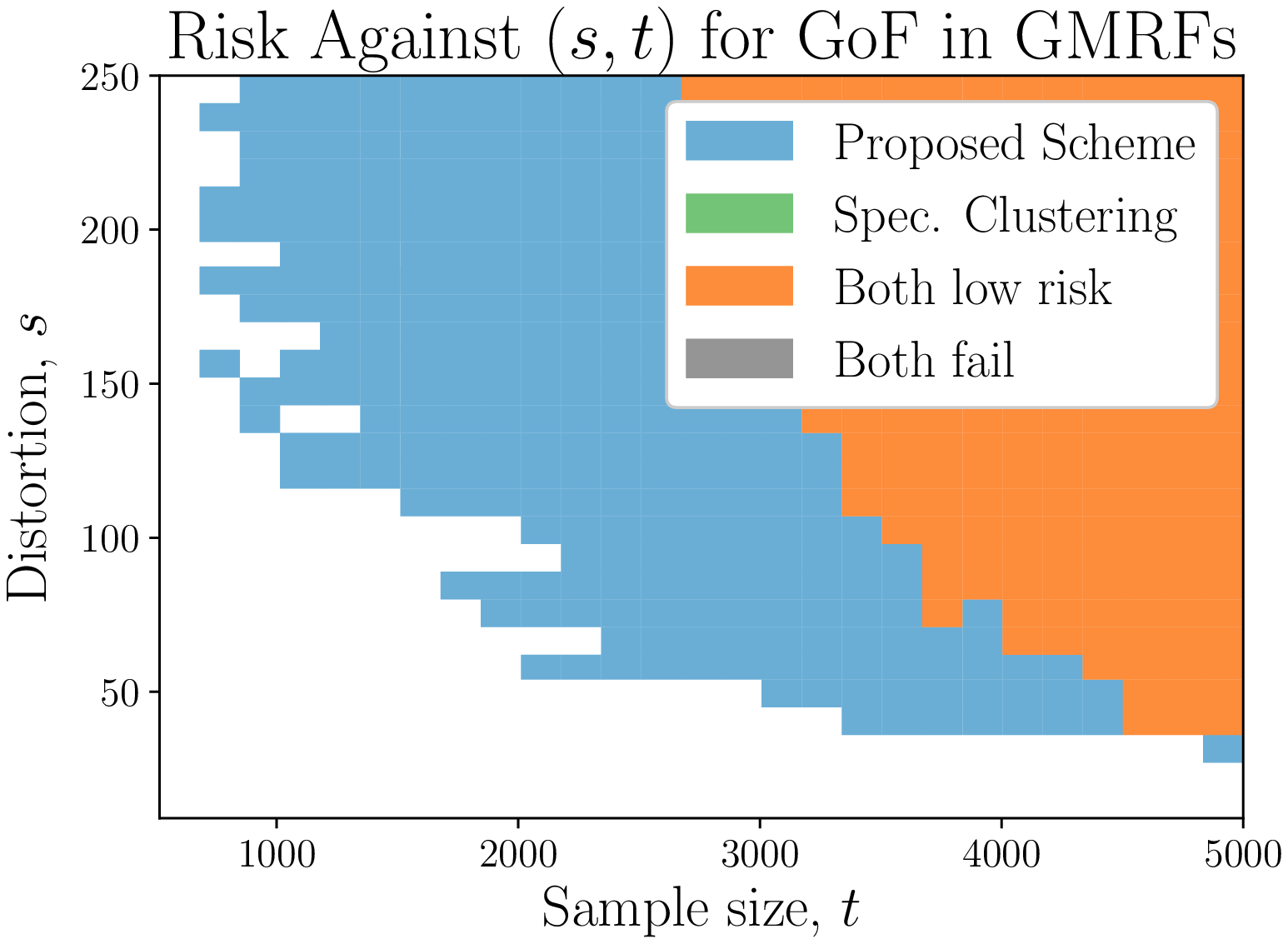}
    ~
            \includegraphics[width=0.45\textwidth]{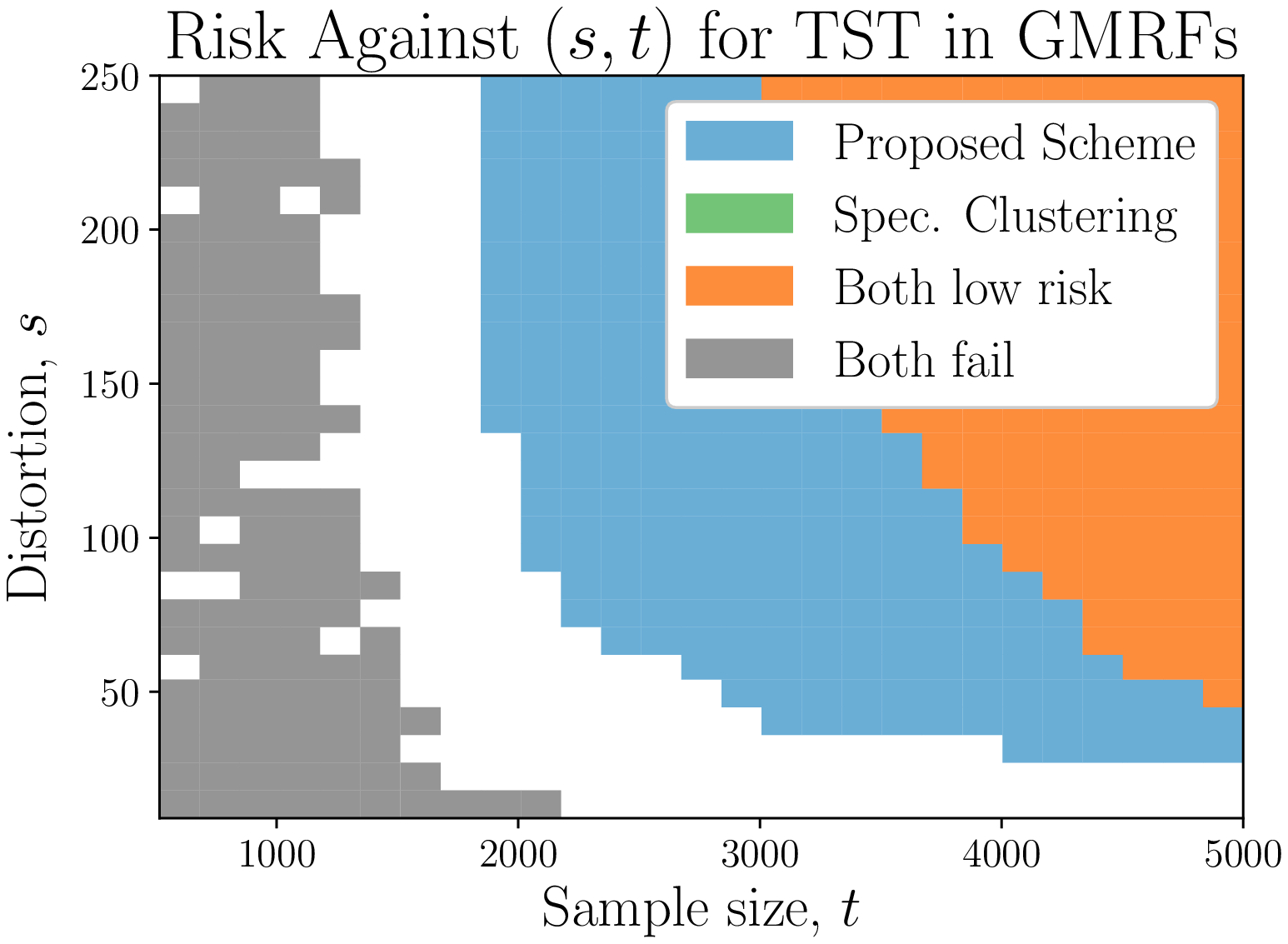}
    \caption{Risks for adaptation of our tests to GMRFs - colour scheme is retained from Fig. \ref{fig:sbm}. The plots show structural similarity to Fig. \ref{fig:sbm}, but with two differences - In GoF, we don't find a high risk region at the sample sizes considered, and the proposed scheme always outperforms the Na\"{i}ve scheme based on spectral clustering.}
    \label{fig:gmrf}
\end{figure}

Following the above heuristic, we na\"{i}vely adapt community recovery and testing to this setting, by replacing all instances of the graph adjacency matrix in previous settings with the sample covariance matrix. Figure~\ref{fig:gmrf} presents our simulations of the risk of this test when $n = 1000$, and $(a,b) \approx (12.3 \log n, 1.23 \log n)$, at $\Lambda \approx 9\log(n)$ (for details see Appx. \ref{appx:exp_gmrf}). This large SNR is chosen so that community recovery would be easy if the graph was recovered;\footnote{Note, however, we expect graph recovery to be impossible at these sample sizes. Lower bounds from \cite{wang2010information} indicate this would require $>3300$ samples theoretically.} this emphasizes the role of the sample size, $t$. Importantly, in this implementation, the threshold for rejecting the null has been fit using data (unlike in the previous sections). This is since we lack a rigorous theoretical understanding of this problem, and have not analytically derived expressions for the thresholds. As a result, these plots should be treated as speculative research intended to underscore the presence of interesting testing effects in this scenario, and to encourage future work along these lines.

\section{{Directions for Future Work}}

The development of the recovery problem for SBMs suggests a number of directions for further work on the testing problems considered above. For instance, one may investigate the exact constants in the testing threshold that the above work suggests, or one may study the testing problem for SBMs with $k > 2$ communities, which is a practically relevant setting since many real-world networks are significantly better described as $k$-SBMs than as $2$-SBMs. In the latter vein, testing problems such as the above may be studied in richer random graph models, such as degree corrected SBMs, or geometric block models. Additionally, testing of strongly imbalanced communities, where one of the communities has size sublinear in $n$ is conceptually unexplored and of interest.

One open problem that draws from the above exposition is if there exists an algorithm for TST in the 2 community setting that does not pass through a partial recovery step and yet works for sparse graphs. We expect that such a method would be necessary for determining exact testing thresholds (for large changes), since the recovery step neccessarily requires some subsampling, which reduces the effective SNR available for testing. In addition, this would be conceptually pleasant, and would eliminate the dissonance in the above work where showing testing guarantees requires passing through recovery guarantees. Such a scheme would also more generally allow study of the testing problem for situations where partial recovery is ill understood.

Finally, we mention that more work is needed on the practical investigation of the effectiveness of the above methods - while the experiments we have run validate the theory, the real-world applicability of the methods above require deeper experimentation. A significant lacuna for this line is the lack of a good real-world dataset for the testing of commumity changes.

\subsubsection*{Acknowledgement} We thank the anonymous reviewers of this and of previous versions of this paper, particularly one for suggesting the proof of the converse in Theorem \ref{thm:tst}.

\printbibliography

\begin{appendix} 
\section{Proofs omitted in Section \ref{sec:gof} }

\subsection{Proof of Achievability in Theorem \ref{thm:gof_bal}}\label{appx:gof_bal_ach}

\begin{proof}[\hspace{-6pt}\nopunct]

We will restrict attention to the case $a > b$ below. The $b > a$ case follows identically. Recall the test in this setting: \[ N_a^{x_0}(G) \overset{H_1}{ \underset{H_0} \gtrless} \frac{bn}{4} + C_1 \max( \sqrt{nb \log(2/\delta)}, \log(2/\delta) ),\] where $C_1$ is the constant implicit in Lemma \ref{lem:bal_gof_guts} below.

Under the null distribution, $N_a^{x_0} (G)$ is distributed as $\mathrm{Bin}(n^2/4, b/n),$ while under the alternate, it is distributed as $\mathrm{Bin}( (n-s)^2/4 + s^2/4, b/n) * \mathrm{Bin}( s(n-s)/2, a/n).$ These distributions can be separated by Bernstein concentration bounds \cite[][Ch. 2]{Chung:2006:CGN:1208774}, as summarised by the following Lemma, which is proved in subsequent sections.
\begin{mylem}\label{lem:bal_gof_guts} There exist constants $C_0,C_1 >1$ such that, if $nb + s(a-b) > C_0 \log(1/\delta),$ then with probability at least $1 - \delta/2$
\begin{enumerate}
	\item[($\alpha$)] Under $H_0$:  \( \displaystyle  N_a^{x_0}(G) \le \frac{bn}{4} + C_1\max\left(\sqrt{nb \log(2/\delta)}, \log(2/\delta)\right).\) 
	\item[($\beta$)] Under $H_1$: \(\displaystyle N_a^{x_0}(G) \ge \frac{bn}{4} + \frac{s (a-b)}{4} - C_1 \sqrt{ (nb + s(a-b))\log(2/\delta) }.\)
\end{enumerate}
\end{mylem}

As the proof of the above lemma discusses, results of the above type hold in the more generic situation where both the communities and the changes can be unbalanced, so long as each community is of at least linear in $n$ size. This allows one to extend the entirety of this theorem to the setting $n^+n^- = \Omega(n^2)$ on replacing $bn/4$ above with $\mathbb{E}_{\mathrm{Null}}[N_a^{x_0}(G)],$ where $n^+$ and $n^-$ are the sizes of the two communities, i.e., the number of $i$ such that $x_i = +1$ and $x_i = -1$ respectively.

Since $s|a-b| \ge s\Lambda \ge C \log(2/\delta),$ the lemma above holds in our setting on picking $C$ large enough. $(\alpha)$ in Lemma \ref{lem:bal_gof_guts} indicates that the false alarm error of test is $\le \delta/2$. Further, since $(nb + s(a-b)) \log(2/\delta) > \log^2(2/\delta),$ part $(\beta)$ shows that missed detection error is $\le \delta/2$ if \begin{align*}
\frac{1}{4} s(a-b) > 2C_1 \sqrt{ (nb + s(a-b) )\log (2/\delta)} \iff \frac{(a-b)^2}{nb + s(a-b)} > C\frac{\log(2/\delta)}{s^2}. \end{align*}

The argument is concluded by some casework:\begin{enumerate}[label = (\roman*), leftmargin = 15pt, topsep = 1pt, itemsep = 2pt]
    \item If $nb \le s(a-b),$ then the left hand side of the condition above can be bounded from below by $s(a-b)/2,$ and thus $s (a-b) \ge 2C_1\log(2/\delta)$ is sufficient. But $s(a-b) \ge s(a-b)^2/(a+b) = s\Lambda$ is larger than $C\log(1/\delta),$ and choosing $C$ large enough is sufficient.
    \item On the other hand, if $nb > s(a-b)$, the left hand side is instead lower bounded by $s^2(a-b)^2/2nb \ge {s^2\Lambda/2n},$ and thus $s^2\Lambda \gtrsim n\log(2/\delta)$ is sufficient to satisfy the same. \qedhere  
\end{enumerate}  
\end{proof}

\subsubsection{Proof of Lemma \ref{lem:bal_gof_guts}}\label{appx:gof_bal_ach_lem}

The proof proceeds by establishing the centres of the statistic $N_a^{x_0}$ under the null and alternate distributions, and then invoking Bernstein-type bounds \cite[][Ch 2]{Chung:2006:CGN:1208774} to show the claimed statements separately.
\begin{enumerate}[leftmargin = 20pt, itemsep = 0pt, topsep = 0pt]
    \item[$(\alpha)$] For the null, $N_a^{x_0}(G)$ is distributed as $\mathrm{Bin}(n^2/4, b/n).$ Thus, clearly $\mathbb{E}_{\mathrm{Null}}[N] = bn/4$. Further, by Bernstein's inequality for the upper tail, \begin{align*} P_{\mathrm{Null}}(N_a^{x_0}(G) > \mathbb{E}_{\mathrm{Null}}[N_a^{x_0}(G)] + nt) &\le \exp{ - \frac{ n^2/4\times  t^2/2}{n^2/4 \times (b/n)+ nt/3}}\\
                                                &\le \exp{ -\frac{3}{2} \frac{nt^2}{b + 4t} } \le \exp{-\frac{3}{8} \frac{nt^2}{t + b} }.
    \end{align*}
    Thus, if \[ \frac{nt^2}{b + t} \ge \frac{8}{3}\log(2/\delta),\] then this tail has mass at most $\delta/2.$ We may now consider the two cases \begin{enumerate}
        \item[(i)] If $nb \le 16/3 \log(2/\delta)$, then plugging in $t = 16/3 \frac{\log(2/\delta)}{n}$ above yields that the the condition is satisfied, since then \[ \frac{nt^2}{b + t} \ge \frac{nt^2}{2t} = \frac{nt}{2} = \frac{8}{3}\log(2/\delta).\]
        \item[(ii)] If $nb \ge 16/3 \log(2/\delta),$ then setting $t =  \sqrt{(16/3) \frac{b}{n} \log 2/\delta}$ we can bound \[ \frac{nt^2}{b + t} = \frac{16/3 \log(2/\delta)}{1 + \sqrt{(16/3)\log(2/\delta)/nb}} \ge \frac{16/3 \log(2/\delta)}{2}.\]
    \end{enumerate}
    As a consequence, picking $nt = \max(\sqrt{ (16/3) nb \log(2/\delta)}, 16/3 \log(2/\delta))$ implies that the probability in question is at most $\delta/2.$
    
    We note that this calculation can be made more robust, in that if the communities are unbalanced but linearly sized with $n$, then the number of edges crossing is $n^+ (n - n^+) = \Omega(n^2)$ in the above, and essentially the same goes through with $n^2/4$ replaced by $n^2/C$ for some constant $C$.
    
    \item[$(\beta)$] This proof proceeds in much the same way as the above. With the modification that the distribution of $N_a^{x_0}(G)$ is now $\mathrm{Bin}( n^2/4 - s(n-s)/2, b/n) * \mathrm{Bin}(s(n-s)/2, a/n),$ since $2\times s(n-s)/4$ of the edges are now between nodes of the same communities. The centre of this is easily seen to be $\frac{nb}{4} + \frac{s(n-s)}{2} \frac{a -b}{n}.$ Further invoking the Bernstein lower tail, we find that  \begin{align*} P_{\mathrm{Alt}}( N_a^{x_0}(G) \le \mathbb{E}_{\mathrm{Alt}}[N_a^{x_0}(G)] -nt) &\le \exp{ -\frac{1}{2} \frac{n^2t^2}{ \frac{s(n-s)}{2} \cdot \frac{a}{n} + \frac{n^2 -2s(n-s)}{4} \cdot \frac{b}{n}}} \\ 
    &\le \exp{ - \frac{n^2t^2}{n b + s(a-b)}} \end{align*} 
    
    The required claim now follows directly by setting $t = \sqrt{\frac{ (nb + s(a-b)) \log(2/\delta)}{n}}$. 
    
    Again, the above can also be rendered more robust to imbalance. Suppose that the communities and the changes are both imbalanced, and let $n^+, n^-$ be the sizes of the communities in $x_0,$ and $s^+, s^-$ be the number of nodes that are moved from $+$ to $-$ and vice-versa according to the alternate $x$. Then the number of edges which behave according to $a/n$ in the alternate is $\tau = s^+(n^- - s^-) + s^-(n^+ - s^+) . $ But $\tau \le s ( n^+ + n^- - s^+ - s^-) = sn,$ so the concentration results go through with a weakening of a factor of $2$. Further, assume wlog that $s^+ \ge s^-.$ since $s^+ + s^- = s,$ and $n^+ + n^- = n,$ we have that \[ \tau = s^+ (n - s) + (s- 2s^+)(n^+ - s^+).\] Minimising the above subject to $s^+ \in [s/2:s],$ we find that the minima can be uniformly lower bounded by $\min( s \min(n^+, n^-), s(n-s)/2)$. So long as each community is of linear size, this is $\Omega(sn),$ and thus the centre of the statistic moves by $\Omega( s(a-b))$ with respect to the null statistic. 
    
    Putting the two effects above together, we can write that under the alternate distribution, with probability $\ge 1- \delta/2,$ \[ N_a^{x_0}(G) \ge \mathbb{E}_{\mathrm{Null}}[N_a^{x_0}(G)] + \frac{1}{C_1} s(a-b) - C_2\sqrt{ (nb + s(a-b)) \log(2/\delta)}.\] 
    
    In conjunction with the discussion for unbalanced but linearly sized communities in case $(\alpha),$ the above allows the claims of the achievability part of Theorem \ref{thm:gof_bal} to hold for the case where both communities are of linear size and changes are not constrained to be balanced without any change other than a weakening of the constants implicit in the same. The only modification required for this is to update the tests to threshold at $\mathbb{E}_{\mathrm{Null}}[N_a^{x_0}(G)] + (\textrm{fluctuation term})$ instead of at $bn/4$ as presented in the main text.
\end{enumerate}

    \subsection{Proofs of converse bounds from Theorem \ref{thm:gof_bal}} \label{appx:gof_conv}
    
    This section begins with an exposition of Le Cam's method, which is the general proof strategy we employ to show both these converse bounds. This is followed by separate subsections devoted to each converse bound claimed in Theorem \ref{thm:gof_bal}.
    
    \subsubsection{Le Cam's method.}\label{sec:lecam}
    
    The generic lower bound strategy is constructed by noting that the minimax risk of the goodness-of-fit problem is lower bounded by the risk of the same with any given prior on the alternate communities, i.e.~ the risk of the problem \[ H_0: x = x_0 \quad \textrm{vs} \quad H_1: x \sim \pi \] for a $\pi$ supported on $\{x: d(x,x_0) \ge s\}$ (or some restriction of the same, as in the following sections), and the Bayes risk \[  R_\pi := \inf_{\varphi: G \to \{H_0, H_1\}}  P( \varphi = H_1 | x_0) + \sum_{x : d(x, x_0) \ge s} \pi(x) P(\varphi = H_0| x).\] By classical Neyman-Pearson theory \cite[see, e.g.,][]{lehmann2006testing}, the likelihood ratio test is optimal under the above risk, and \[ R_{\pi} = 1 - d_{\mathrm{TV}}\left(P(G|x_0) , \ip{P(G|x)}_\pi \right), \] where $\ip{P(G|x)}_{\pi} := \sum_x \pi(x) P(G|x)$, and $d_{\mathrm{TV}}$ is the total variation distance \[ d_{\mathrm{TV}}(P,Q) := \frac{1}{2} \|P - Q\|_1.\]
    
    We proceed by bounding $d_{\mathrm{TV}}$ by an $f$-divergence more conducive to tensorisation in order to exploit the (conditional) independence of the edges in an SBM, and then by choosing an appropriate $\pi$. The $f$-divergence inequalities we use are \begin{enumerate}
        \item $\chi^2$ bound: Recall that \[ D_{\chi^2}(Q\|P) = \sum_x P(x) \left( \frac{Q(x) - P(x)}{P(x)}\right)^2  = \mathbb{E}_P[ L^2(X)] -1, \] where $L(x) := Q(x)/P(x)$ is the likelihood ratio. It holds that \[ d_{\mathrm{TV}}(P,Q) \le \sqrt{\frac{1}{2}\log( 1 + D_{\chi^2}(Q\|P) )},\] which follows from Pinsker's inequality and the fact that \[ D_{\mathrm{KL}}(Q\|P) \le \log(1 + D_{\chi^2}(Q\|P)),\] which is a consequence of Jensen's inequality applied to the $\log$ (or, equivalently, the monotonicity of R\'{e}nyi divergences).
        
        Invoking the above inequality and Le Cam's method, we find that for any choice of $\pi,$ and for $L(G) := \frac{\ip{P(G|x)}_\pi}{P(G|x_0)},$ the following is necessary for the minimax risk of the GoF problem to be bounded above by $\delta:$ \[ \mathbb{E}_{x_0}[L^2(G)] \ge \exp{2(1-\delta)^2}. \] For $\delta \le 1/4,$ this yields a necessary lower bound of $\mathbb{E}_{x_0}[L^2] > 3.08.$
        
        \item Hellinger bound: The Bhattacharya coefficient of $P,Q$ is defined as \[ \mathrm{BC}(P,Q) := \sum_x \sqrt{P(x) Q(x)},\] and the Hellinger divergence as \[ D_{\mathrm{H}}(P,Q) := \sqrt{1 - \mathrm{BC}(P,Q)} = \frac{1}{\sqrt{2}} \|\sqrt{P} - \sqrt{Q}\|_2.\] We exploit the relation \[ d_{\mathrm{TV}}(P,Q) \le \sqrt{D_{\mathrm{H}}^2(P,Q)(2-D_{\mathrm{H}}^2(P,Q) )} = \sqrt{1 - \mathrm{BC}^2(P,Q)}, \] which is a consequence of the Cauchy-Schwarz inequality. 
        
        Again plugging this in with $Q = \ip{P(G|x)}_\pi,$ we find that in order for the risk to be smaller than $\delta,$ we must have that \begin{align*}
            \delta &\ge  1- \sqrt{1 - \mathrm{BC}^2} \ge \frac{\mathrm{BC}^2}{2} \implies
            \mathrm{BC} \le \sqrt{2\delta},
        \end{align*}
        where $\mathrm{BC} = \mathrm{BC}(\ip{ P(G|x)}_\pi, P(G|x_0))$.
    \end{enumerate}
    
    We now proceed to show the claimed bounds. Recall that we are required to show that if $R_{\mathrm{GoF}} < \delta \le 1/4,$ them \begin{align}
        \Lambda &\gtrsim \log(1 + n/s^2) \label{gof_lowb_1} \\
        s\Lambda &\gtrsim \log(1/\delta). \label{gof_lowb_2}
    \end{align}

    \subsubsection{Proof of the converse bound (\ref{gof_lowb_1}) }\label{appx:gof_conv_mu_pf}
        
    For convenience, we let \begin{equation} \nu := (a-b)^2 \left( \frac{1}{a (1-a/n)} + \frac{1}{b(1-b/n)} \right). \end{equation}  Since $a,b \le n/2,$ and since $a/b = \Theta(1),$ we have $\Lambda \asymp \nu,$ and it suffices to show the same bound on the latter. 

    We invoke Le Cam's method with a $\chi^2$-bound. Let $m:= n/2, t:= s/2 $ and let $x_0$ be the partition $([1:m], [m+1:2m]).$ 

    The alternate prior is chosen to be the uniform prior on the set of alternate partitions constructed as follows. For each $T \subset [1:m],$ we define the partition \begin{align*}y_T(+) &= [1:m] \cup (m + T) \sim T \\ y_T(-) &= [m+1:2m] \cup T \sim (m+T),\end{align*} where $(m+T) = \{ i + m : i \in T\}.$ Let $\mathcal{Y}_t := \{ y_T: T \subset [1:m], |T| = t\}.$ For conciseness, we define the measures on $\mathcal{G}:$ \[ P_{y_T}(\cdot) := P_T(\cdot) :=  P(\cdot \mid y_T),\] and set $P_0 = P(\cdot\mid x_0).$ Further, for convenience, we set $p = a/n$ and $q = b/n.$

    For a graph $G,$ we find that $L(G) :=  \frac{1}{|\mathcal{Y}_t|} \sum_{x \in \mathcal{Y}_t}\frac{P_{x}(G)}{P_0(G)}.$ To invoke Le Cam's method (\S\ref{sec:lecam}), we need to upper bound $\mathbb{E}_{P_0}[L^2(G)].$

    To this end, we will define for an edge $e = (u,v)$, and a graph $G$ (which is implicit in the notation) \begin{equation}  f_{e}(q,p) := (q/p)^{e} ( (1-q)/(1-p) )^{1- e}. \end{equation} Above, $f_e(q,p)$ arises as a ratio of the probabilities of a $\textrm{Bern}(q)$ and a $\textrm{Bern}(p)$ random variable. Thus, it is the likelihood ratio of an edge being between nodes in the different and in the same community. 

    First observe that \begin{equation}  \frac{P_{T}}{P_0} =   \left(~~~~\prod_{ \mathclap{\substack{i \in [1:m] \sim T,\\ j \in m + T }}} f_{ij}(p,q) \right) \left(~~~~~~~~~~\prod_{ \mathclap{\substack{i \in [m + 1:2m] \sim m+ T,\\ j \in  T }}} f_{ij}(p,q)  \right)  \left(~~~~\prod_{ \mathclap{\substack{i \in [1:m] \sim T,\\ j \in T }}} f_{ij}(q,p) \right)  \left(~~~~~~~~~~\prod_{\mathclap{ \substack{i \in [m+ 1:2m] \sim m + T,\\ j \in m+  T }}}  f_{ij}(q,p)   \right)\end{equation}

    An important feature of the setup above is that every term in the above product is independently distributed, and wherever $f_{ij}(p,q)$ appears, the corresponding $e_{ij}$ is $\textrm{Bern}(q)$, and similarly with $f_{ij}(q,p)$ and $\textrm{Bern}(p)$.

\begin{figure}[ht]
    \centering
        \includegraphics[width=.85\linewidth]{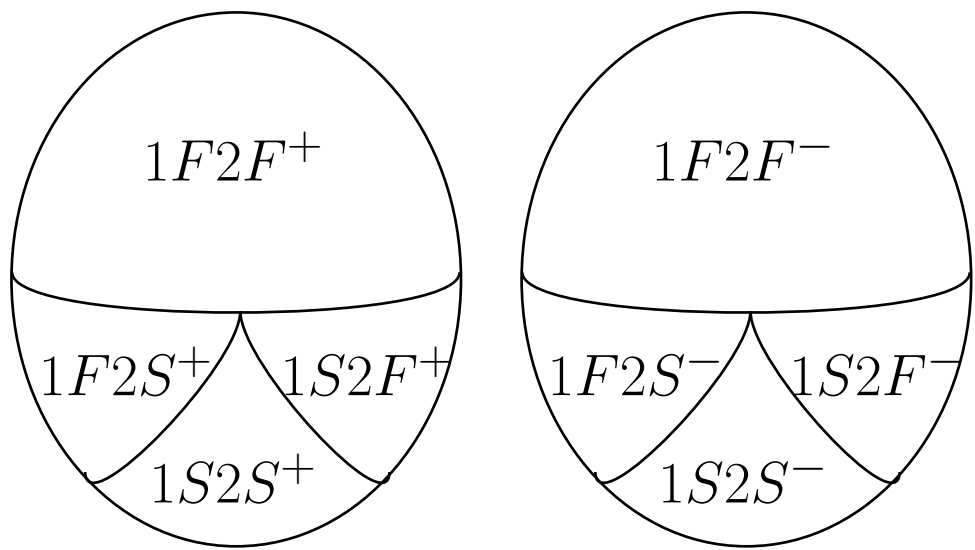}
        \caption{A schematic of the nodes, partitioned according to their labellings in $x_0, y_{T_1}, y_{T_2}$. The two ovals denote the partition induced by $x_0$ into groups marked $+$ and $-$. The section $1F2F^+$ denotes the nodes in the $+$ group whose labels remain \emph{fixed} to $+$ in both $y_{T_1}, y_{T_2}.$ The section marked $1S2F^+$ denotes the nodes in the $+$ group whose labels are \emph{switched} to $-$ in $y_{T_1}$ but remain \emph{fixed} to $+$ in $y_{T_2}$. Other labels are analogously defined.}
    \label{fig:goft_lowb}
    \end{figure}

    Note that \[ \mathbb{E}_{P_0} [L^2(G)] = \sum_{T_1, T_2 \subseteq [1:m] \textrm{ of size $t$}} \mathbb{E}_{P_0} \left[ \frac{P_{T_1}(G) P_{T_2}(G)}{P_0^2(G)} \right], \] and so we must control expectations of this form in order to apply Le Cam's method. Let us fix $T_1$ and $T_2$ for now, and partition the nodes into groups as described by the Figure \ref{fig:goft_lowb}\footnote{The argument, while simple, gets a little notationally hairy at this point. We recommend that the reader consults Figure \ref{fig:goft_lowb} frequently, preferably a printed copy that allows one to sketch the various types of connections on it.}.\\

    Note that in the figure, $1F2F^+ = [1:m] \sim (T_1 \cup T_2), 1S2S^- = (m+T_1) \cap (m+t_2)$ and so on. Also, importantly, the size of groups with the same number of $S$s and $F$s in the above representation is identical (i.e., $|1F2S^+| = |1F2S^-| = |1S2F^+| = |1S2F^-|$ and so on.)

    We consider how the terms relating to the edge $(u,v)$ for any $u,v \in [1:2m]$ appear in the product $\frac{P_{T_1}P_{T_2}}{P_0^2}.$ Below, 
    \begin{itemize}
        \item Clearly, if $u$ and $v$ are both in the same group in both settings, the behaviour of the edge $(u,v)$ under the alternate distributions and the null distribution is identical, and these terms will not appear in the product.
        \item If both $(u,v) \in 1F2F^+ \times 1F2F^- \cup 1S2S^+ \times 1S2S^-,$ then again, the edge $(u,v)$ has identical distribution under both alternates and the null, and these terms do not appear in the product.
        \item If $(u,v) \in 1F2F^+ \times 1F2S^+,$ then the $(u,v)$ term does not appear in $P_{T_1}/P_0,$ but appears once in $P_{T_2}/P_0.$ Since likelihoods must average to $1,$ and since the distributions of the edges are independent, any term which appears just once is averaged out when we take expectations with respect to $P_0.$ Thus, even though these terms appear in the product, we may ignore them due to our eventual use of the expectation operator. A quick check will show that the same effect happens for $(u,v) \in \Gamma_1 \times \Gamma_2,$ where $\Gamma_1$ can be obtained by inverting one instance of an $F$ to a $S$ or vice versa, and possibly changing the sign (e.g. $1F2S^- \times 1S2S^+.$) Thus, all such pairs can be safely ignored.
        \item This leaves us with edges of the form $ \{1F2F^\pm \times 1S2S^\pm\} \cup \{ 1F2S^\pm \times 1S2F^\pm\}.$ In these cases, if the signs of the two choices match - i.e. \[ (u,v) \in \Gamma^+ \times \widetilde{\Gamma}^+ \textrm{ for } ( \Gamma, \widetilde{\Gamma}) \in \{ (1F2F, 1S2S), (1S2F, 1F2S) \},\] then we will obtain a contribution of $f_{uv}(q,p)^2$ to the product. On the other hand, if they differ, then we will obtain a contribution of $f_{uv}(p,q)^2$ 
    \end{itemize}

Accounting for the above, and taking expectation, we have that 
\begin{equation}\label{eqn:gof_giant_eq} \mathbb{E}\left[ \frac{P_{T_1} P_{T_2}}{P_0^2} \right]  =  \left( \Psi \right)^{ |1F2F^+| \cdot |1S2S^+| + |1F2F^-| \cdot |1S2S^-|  + |1S2F^+|\cdot |1F1S^-| + |1F2S^+| \cdot |1S2F^-|} ,\end{equation}

where \begin{equation}
    \Psi := \mathbb{E}_{\textrm{$e \sim $ Bern$(p)$}} [f_{e}(q,p)^2] \mathbb{E}_{\textrm{$e \sim$ Bern$(q)$}} [f_{e}(p,q)^2]
\end{equation}

Further, since in our choice of the alternate communities the groups with the same number of $S$s and $F$s have identical size, and thus we may rewrite (\ref{eqn:gof_giant_eq}) above as \[ \mathbb{E}\left[ \frac{P_{T_1} P_{T_2}}{P_0^2} \right] = \Psi^{2 (|1F2F^+||1S2S^+| + |1S2F^+|^2)}. \]

For convenience, let $|1S2S^+| = |T_1 \cap T_2| = k.$ We then have that $|1S2F^+| = t - k$ and $|1F2F^+| = m + k - 2t.$

We thus have that \begin{align} \mathbb{E}_{P_0} \frac{P_{T_1} P_{T_2}}{P_0^2}  &= \exp{ (\log \Psi) ( 2 k(m+k - 2t) + 2 (t-k)^2 )}\\ &= \exp{ (\log \Psi) (2mk + 2k^2 - 4kt + 2k^2 +2 t^2  - 4kt)} \\ &= \exp{ (\log \Psi) (2mk + 4k^2 - 8kt + 2t^2)}\\ &\le \exp{ (\log \Psi) ((2m - 4t)k + 2t^2 )},\end{align}

where we have used that $k \le t.$

Now, for $(p,q) = (a/n, b/n),$  \begin{align}
    \Psi &= \left( \frac{q^2}{p} + \frac{(1-q)^2}{(1-p)}\right)\left( \frac{p^2}{q} + \frac{(1-p)^2}{(1-q)}\right) \\
      &= \left(1 + \frac{(p-q)^2}{p(1-p)} \right)\left(1 + \frac{(p-q)^2}{q(1-q)} \right) \\
      &= \left(1 + \frac{(a-b)^2}{n a(1-a/n)} \right)\left(1 + \frac{(a-b)^2}{n b(1-b/n)} \right) \\
      &= 1 + \frac{\nu}{n} + O(n^{-2}) \le 1 + 2\frac{\nu}{n} .
\end{align}

As a consequence, using $2m = n,$ and the development above, 

\begin{align} \mathbb{E}_{P_0} \frac{P_{T_1} P_{T_2}}{P_0^2} \le \exp{\frac{4t^2}{n} \nu} \exp{ 2k\nu( 1 - 4t/n)}.\end{align}

The above is insular to the precise identities of $T_1, T_2.$ Further, for a given $T_1,$ the number of partitions $T_2$ such that $|T_1 \cap T_2| = t$ is $\binom{t}{k} \binom{m-t}{t-k}.$ Feeding this into the expression for $\mathbb{E}[L^2(G)]$ and some simple manipulations yield that \begin{equation} \label{ineq:likelihood} \mathbb{E}_{P_0}[L^2(G)]  \le \frac{e^{\frac{4t^2}{n} \nu} }{\binom{m}{t}} \sum_{k = 0}^t \binom{t}{k} \binom{m-t}{t-k} \exp{ 2k \nu(1 - 4t/n)},\end{equation}

where we remind the reader that $t = s/2, m = n/2.$ 

Recall from \S\ref{sec:lecam} that if $\mathbb{E}_{P_0}[L^2] < 3,$ then the risk exceeds $0.25.$ Thus, we will aim to upper bound (\ref{ineq:likelihood}) by $3$.
    
    {We begin by rewriting}
    {
    \begin{align} \mathbb{E}_{P_0}[L^2(G)]  &\le \frac{e^{\frac{4t^2}{n} \nu} }{\binom{m}{t}} \sum_{k = 0}^t \binom{t}{k} \binom{m-t}{t-k} \exp{ 2k \nu(1 - 4t/n)}, \\ 
    &= e^{\frac{4t^2}{n} \nu} \mathbb{E}[\xi^Z],\end{align}
    }
    {where $\xi := \exp{2 \nu (1- 4t/n)} > 1$ and $Z = \sum_{i = 1}^t Z_i,$ where $Z_i$ are sampled without replacement from the collection of $t$ (+1)s and $m-t$ (0)s. Note that $z \mapsto \xi^z$ is continuous and convex for $\xi \ge 1.$ By Theorem 4 of \cite{Hoeffding}, \[ \mathbb{E}[ \xi^Z] \le \mathbb{E}[ \xi^{\widetilde{Z}}],\] for $\widetilde{Z}= \sum_{i = 1}^t \widetilde{Z}_i,$ where $\widetilde{Z}_i$ are drawn by sampling with replacement from the same collection. But $\widetilde{Z}$ is just a Binomial random variable with parameters $(t, t/m).$ Thus, we have that }
    
    {\begin{align} \label{ineq:likelihood_2} \mathbb{E}_{P_0}[L^2(G)]  &\le {e^{\frac{2t^2}{m} \nu}} \left(1 + \frac{t}{m} \left( \exp{2 \nu (1- 2t/m)} - 1 \right)  \right)^t \\
                                            &\le \exp{2\frac{t^2}{m}\nu + \frac{t^2}{m} \left( \exp{2 \nu (1- 2t/m)} - 1 \right)} \\
                                            &\le \exp{\frac{t^2}{m} \left( 2\nu + e^{2\nu} - 1\right)} \\
                                            &\le \exp{ 2\frac{t^2}{m} \left(  e^{2\nu} - 1\right)},\end{align} }
    
    {where the final inequality uses $ u  < e^{u} - 1$. Using the above, and noting that $m/2t^2 = n/s^2,$ we find that \[ \nu \le \frac{1}{2} \log\left(1 + \frac{\log(3) n}{s^2} \right) \implies \mathbb{E}_{P_0}[L^2(G)] \le 3, \]} finishing the argument. \hfill $\qedee$
    
    \subsubsection{Proof of the converse bound (\ref{gof_lowb_2})}\label{appx:gof_bal_conv_lam_pf}
    
    Recall that this part of the theorem claims that if $R_{\mathrm{Gof}} \le \delta \le 1/4,$ then $s\Lambda \ge C \log(1/\delta).$

    We will again use Le Cam's method (\S\ref{sec:lecam}), this time controlling the total variation distance by a Hellinger bound. 

    Let $x_0 = ([1:n/2] , [n/2 + 1: n])$ be the null partition, and $\mathcal{Y}:= \{ y\},$ with $y:= ([1:n/2 - s/2] \cup [n/2+1:n/2 + s/2] , [n/2-s/2+1:n/2] \cup [n/2+ s/2 + 1: n])$. We let $P_{x_0}(G) := P(G|x_0), $ and similarly $P_y.$ Recall from the section on Le Cam's method that the following is a necessary condition for the risk to be smaller than $\delta$ \[ \mathrm{BC}(P_{x_0}, P_y) \le {\sqrt{2\delta}}. \] The Bhattacharya Coefficient can be estimated directly in this setting. (We omit the derivation below)    \begin{align} \mathrm{BC}(P_{y}, P_{x_0}) &= \left( \sqrt{\frac{ab}{n^2}} + \sqrt{\left(1-\frac{a}{n}\right)\left(1-\frac{b}{n}\right)}\right)^{s(n-s)} \end{align}
    
    For $ u, v < 3/4,$ \[ \sqrt{(1-u)(1-v)} \ge 1 - (u+v)/2 - 2(u - v)^2 . \]

    Thus \begin{align} \mathrm{BC}(P_{y}, P_{x_0}) &\ge \left(1 - \frac{a + b}{2n} + \frac{ab}{n} - 2\frac{(a-b)^2}{n^2}\right)^{s(n-s)} \\ &= \left( 1 - \frac{(\sqrt{a} - \sqrt{b})^2}{2n} - 2\frac{(a-b)^2}{n^2} \right)^{s(n-s)} \\ &\ge \left( 1 - \frac{(\sqrt{a} - \sqrt{b})^2}{n} \right)^{s(n-s)} \\  &\ge \exp{- 2s (\sqrt{a} - \sqrt{b})^2 },\end{align}
    
    where the third inequality uses $(a+b) < n/4$, and the final uses used $1-u \ge e^{-2 u}$ for $0 < u \le 0.75$---which applies since $0 <  (\sqrt{a} - \sqrt{b})^2 < \max(a,b) < n/4$---and $n-s \le n.$ 
    
    Now note that \[  (\sqrt{a} - \sqrt{b})^2 = \frac{(a-b)^2}{(\sqrt{a} + \sqrt{b})^2} \le \frac{(a-b)^2}{a + b} = \Lambda,\] and thus, \[  \mathrm{BC}(P_{y}, P_{x_0}) \ge \exp{-2 s \Lambda}.\]
    
    Invoking the condition for $R_{\mathrm{GoF}} \le \delta$ above, we have \begin{align*}
        \exp{- 2s \Lambda } \le \sqrt{2\delta}& \\
        \iff s \Lambda \ge \frac{1}{4} \log \frac{1}{2\delta}&.
    \end{align*}
For $\delta \le 1/4,$ we may further lower bound the above by $(\log(1/\delta))/8.$ \hfill $\qedee$

\subsection[A comment on the SNR when the ratio of intra- and inter- community degrees is not a constant]{ A comment on the role of $\Lambda$ when $a/b \neq \Theta(1)$ }

The main text concentrates on the setting where $a/b$ is a constant. Here, we briefly comment on the setting where the ratio $\rho:= \frac{\max(a,b)}{\min(a,b)}$ is \emph{diverging} with $n$. In the setting of balanced communities and divergent $\rho$, the behaviour of the goodness-of-fit problem is no longer described by the quantity $\Lambda = \frac{(a-b)^2}{a+b},$ but instead depends on \[ \mu:= \frac{(a-b)^2}{\min(a,b)} .\] Specifically, our proofs can, with minimal changes, be adapted to say that for balanced GoF, $R_{\mathrm{GoF}}$ can be solved with vanishing risk if the following hold: \begin{align*}
    s\Lambda &= \omega(1) \\
    \mu &= \omega( n/s^2), 
\end{align*}

and further, to attain the same, it is necessary to have \begin{align*}
    s\Lambda &= \omega(1) \\
    \mu &\gtrsim \log(1 + n/s^2).
\end{align*}

Indeed, for the lower bounds, $\mu \le \nu \le 4\mu$ uniformly, where $\nu$ is the SNR quantity in the previous section, and the upper bounds naturally feature $\mu$.

Together, the above offer a tight characterisation of the GoF problem in the setting of balanced communities and \emph{large} $s$. Note that $\nicefrac{\mu}{\Lambda} = 1 + \rho$ diverges with $\rho$, and thus the above indicate that GoF testing becomes much easier as this ratio blows up - something to be expected.

Despite the above developments, we concentrated on the setting $\rho = \Theta(1)$ in the main text. This is largely because the majority of the literature on the SBM focuses on this regime, as this is the hardest setting for inference about the planted structure.  Thus, in order to compare to existing work, we examined the $a \asymp b$ setting.  


As an aside, we note that unlike the above GoF results, the TST results do not alter in the setting of divergent $\rho.$ Theorem \ref{thm:tst}, and in particular the converse bound $\Lambda \gtrsim 1,$ continues to hold for this setting. 

On the whole, this line of work is still under investigation, particularly whether the behaviour of GoF for large $\rho$ continues to be driven by $\mu$ in the setting of small changes. We plan to explore this question in later work.

\section{Proofs omitted from section \ref{sec:tst} }\label{appx:tst_proofs}

\subsection{Proof of Achievability in Theorem \ref{thm:tst}}\label{appx:tst_ach}

\begin{proof}[\hspace{-6pt}\nopunct] 
    We carry out the analysis for the case $a > b.$ The reverse can be handled similarly. Note that this assumption will implicitly be made in all the lemmata that follow.
    
    Recall that the scheme in Algorithm~\ref{alg:tst} utilises a partial recovery routine. For the purposes of the following argument, we invoke the method of \cite{chin2015stochastic}, which provides a procedure that, under the conditions of the theorem, that attains with probability at least $1 - 1/n$ recovery with at most $\varepsilon_{\max} n$ errors, where $\varepsilon_{\max} = \min(1/2, 2e^{-C\Lambda})$ for an explicit constant $C$. We choose $\Lambda$ large enough so that $\varepsilon_{\max}$ is bounded strictly below $1/2$ - for convenience, say by $1/3$.
    
    Let $G' \sim P(\cdot | x)$ be an independent copy of $G,$ useful in the analysis, and recall the definition of $\widetilde{G}, G_1$ from Algorithm \ref{alg:tst}. We define the following events that we will condition on in the sequel:  \[ 
    \mathcal{E}(G_1) = \{ \textrm{Number of edges in $G_1$ } \le {an}/{2} \} \qquad \mathcal{E}(\hat{x}) = \{ d(\hat{x}, x) \le \varepsilon_{\max} n \} \]
        For succinctness, we let $\mathcal{E} := \mathcal{E}(G_1) \cap \mathcal{E}(\hat{x}).$
    The analysis proceeds in four steps:  
\begin{enumerate}[label = (L\arabic*), leftmargin = 20pt, topsep = 1pt, itemsep = 2pt]
               \item \begin{shortlem}\label{lem:probS} $P(\mathcal{E}) \ge 1 -4/3n $. \end{shortlem}
    \item \begin{shortlem}\label{lem:tst_null_centre} $\left|\mathbb{E}[ 2T^{\hat{x}}(\widetilde{G}) - T^{\hat{x}}(G') \mid \mathcal{E}]\right| \le a^2.$ \end{shortlem}
    \item \begin{shortlem} If $d(x,y) \ge s,$ then for $\kappa :=  (1-2\varepsilon_{\max})^2 - 1/(n-1)$, \label{lem:tst_integrated}\begin{equation*}
        \mathbb{E}[T^{\hat{x}}(G') - T^{\hat{x}}(H) \mid \mathcal{E}]  \ge \kappa \frac{(a-b)}{n}(n - s)s.
    \end{equation*} 
\end{shortlem}
    \item \begin{shortlem}\label{tst:conc} Let $\xi := a^2 + 5\sqrt{2na\log(6n)}.$ Then \begin{align*} 
    &P_{\mathrm{Null}} \left(T \ge \xi  \middle| \mathcal{E} \right) \le 2/3n  \\ 
    &P_{\mathrm{Alt.}}\left( T \le \kappa (a-b)s/2 - \xi  | \mathcal{E} \right) \le 4/3n \end{align*}
    \end{shortlem}
\end{enumerate}

\noindent We briefly describe the functional roles of the above, and relegate their proofs to the following sections. 
\begin{enumerate}[label = (L\arabic*), leftmargin = 20pt, topsep = 1pt, itemsep = 2pt]
    \item allows us to make use of the typicality of $G_1$ and the recovery guarantees of $\hat{x}$. The former is primarily useful for (L2), while the latter induces (L3).
    \item lets us avoid the technical issues arising from the fact $\widetilde{G}$ and $G_1, \hat{x}$ are correlated, and allows us to work with the simpler $G'.$ It also shows that under the null, the mean of $T$ is small. This lemma is likely loose, and introduces the nuisance condition $a \le n^{1/3}$. 
    \item shows that under the alternate, the centre of $T$ linearly grows with $s$ despite the weak recovery procedure's errors. 
    \item serves to control the fluctuations in $T.$ The $\sqrt{n}$-level term arises from the randomness in $\widetilde{G}, H, G',$ and the $a^2$ term from our use of $G'$ and (L2).
\end{enumerate}

Putting the above together, we find that the risk is bounded by $4/3n + 2/3n + 4/3n \le 4/n$ if \[    
\kappa (a-b)s \ge 4( a^2 + 5\sqrt{2na\log(6n)}).\]

Since $a^2 = a^{3/2}\sqrt{a} \le \sqrt{na},$ and for $\Lambda $ a large enough constant, $\varepsilon_{\max} \le 1/3 \implies \kappa \ge (1/3 - 1/(n-1) )^2 \ge 1/36 $ for $n \ge 7,$ the above condition is equivalent to \[ (a-b)s \ge C' \sqrt{na \log(6n)} \] for a large enough $C'$. Rearranging and squaring, this is equivalent to \[ \frac{(a-b)^2}{a} \gtrsim \frac{n \log(6n)}{s^2}.\] For $s \ge n^{\nicefrac{1}{2} + c}$ as in the statement, the quantity on the right hand side is decaying with $n$. Further, $\Lambda$ is smaller than the left hand side, so it being bigger than a constant forces the above to hold. 

Note that the threshold in Algorithm \ref{alg:tst} alters the fluctuation range above from $\sqrt{na}$ to $\sqrt{n (a+b)}$. The reason for this is that this relaxation allows Algorithm \ref{alg:tst} to be agnostic to the knowledge of $(a,b)$ - generic spectral clustering schemes do not require this knowledge, and the threshold of our scheme depends only on $n(a+b),$ which can be robustly estimated in our setting since the number of edges in the graph is proportional to this. In addition, invoking the bounds of \cite{chin2015stochastic} allows explicit control on $\kappa$ above, and thus provides an explicit value of the constant $C$ in Algorithm \ref{alg:tst}.
\end{proof}

\subsubsection{Relaxing Exact Balance for TST}\label{appx:tst_imbalance}

We briefly discuss the modifications required to the above analysis in order to extend the same to unbalanced but linearly sized communities. Of the four lemmata used in the proof described above, the proof of Lemma $\ref{lem:tst_null_centre}$ is completely agnostic to the sizes of the communities. In addition, while the proof presented in \cite{chin2015stochastic} concentrates on the case of exactly balanced communities, as noted in their Section 1, it can be extended to unbalanced but linearly sized communities with minimal changes, although with a corresponding weakening of the constants in the rate with which error decays with increasing $\Lambda.$ This extends Lemma $\ref{lem:probS}$ to linearly sized communities. 

In contrast, Lemma \ref{lem:tst_integrated} does rely on the assumption of balance. To sidestep this, we show the following version for use in this setting: \begin{mylem}\label{lem:tst_unbal_alt}
Let the communities be of sizes $n_+, n_-.$ If $d(x,y) \ge s,$ then \begin{equation*}
        \mathbb{E}[T^{\hat{x}}(G') - T^{\hat{x}}(H) \mid \mathcal{E}]  \ge \frac{(a-b)s}{2} \left( 1 - 2 \varepsilon \left( \frac{n}{\min(n_+, n_-) } + 2\right)\right).
    \end{equation*} 
    
\end{mylem}

When $\min(n_+, n_-) = c n,$ for some $c>0,$ then by enforcing that $\varepsilon$ is smaller than, say, $\left(3(2+\nicefrac{1}{c})\right)^{-1},$ which may be done by choosing a large enough, but $O(1)$, value of $\Lambda,$ the above can be expressed as $\Omega(s(a-b))$.

Lastly, the alternate case in Lemma \ref{tst:conc} must be adjusted. However, this concentration result is actually proved by arguing that the event $\{T \le \mathbb{E}_{\mathrm{Alt}}[T^{\hat{x}} (G') - T^{\hat{x}}(H)|\mathcal{E}] - \xi\}$ has low probability given $\mathcal{E}$, and the above lemma implies that this expectation is at least $\Omega( s(a-b)) ),$ so the corresponding tail bound goes through to the required form trivially. 

At this point, the concluding remarks of the above proof apply to finish the argument.

\subsection{Proofs of Lemmata used in \ref{appx:tst_ach}}

\subsubsection{Proof of Lemma \ref{lem:probS}} \label{appx:probS}

\begin{proof}[\hspace{-6pt}\nopunct]
  
    We first note that by the work of \cite{chin2015stochastic}, or \cite{fei2019exponential}, under the conditions of the theorem, $\mathcal{E}(\hat{x})$ holds with probability at least $1 - 1/n.$ By a union bound, it suffices to show that $P(\mathcal{E}(G_1) ) \ge 1- \frac{1}{3n}.$
    Recall that \begin{equation} P( (e,v) \in G_1|x) = \begin{cases} \frac{a}{2n} & x_u = x_v \\ \frac{b}{2n} & x_u \neq x_v, \end{cases} \end{equation} and that edges are independent. Thus the number of edges in $G_1$ is a sum of Bernoulli random variables of parameter $\le a/2n.$ The factor of $2$ arises since $G_1$ is sub-sampled at rate $1/2.$ Let $\#G_1$ be the number of edges in $G_1$. We have \begin{align}
        \mathbb{E}[\#G_1&] \le \binom{n}{2} \frac{a}{2n} \le \frac{na}{4} \\
        P( \#G_1 \ge  \mathbb{E}[\#G_1&] + \sqrt{n a \log (3n)} ) \le 1/3n,
    \end{align}
    
    where the first bound follows from inspection, and the second follows from the Bernstein upper tail bound of \cite[][Ch. 2]{Chung:2006:CGN:1208774} and the condition $a \ge 16\log(6n)/n.$ Further invoking this condition we find that $\sqrt{na\log(3n)} \le na/4,$ and thus \begin{equation*}
        P(\mathcal{E}(G_1)) = P(\#G_1 \le na/2) \ge 1 - \frac{1}{3n}.\qedhere 
    \end{equation*}
\end{proof}
\subsubsection{Proof of Lemma \ref{lem:tst_null_centre}} \label{appx:tst_null_centre}
\begin{proof}[\hspace{-6pt}\nopunct]
     Let $$c_{uv} := \frac{(a+b) + (a-b) x_u x_v}{2} \le a.$$ Recall that $c_{uv}/n$ is the probability under $x$ of the edge $(u,v)$ existing.
    
    Also note that for a graph $\Gamma$ and a partition $z,$ \[ T^{z}(\Gamma) = \sum_{1 \le u < v \le n} z_uz_v \Gamma_{uv},\] where $\Gamma_{uv} := \mathbf{1}\{ (u,v) \in \Gamma \}.$
    
    We're interested in controlling \[ T = 2T^{\hat{x}}(\widetilde{G}) - T^{\hat{x}}(G') = \sum \hat{x}_u\hat{x}_v (2\widetilde{G}_{uv} - G'_{uv} ).\] Since $\hat{x}$ is a deterministic function of $G_1,$ $\widetilde{G}$ is independent of $\hat{x}$ given $G_1.$ Further, $G'$ is independent of $(G_1, \widetilde{G}).$ Lastly observe that \[ P( (u,v) \in \widetilde{G} \mid G_1 ) = \frac{{c_{uv}}/{2n}}{1 - c_{uv}/2n} (1 - (G_1)_{uv}). \]
    
    As a consequence, \begin{align}
        \mathbb{E}[T\mid G_1] &= \sum \hat{x}_u \hat{x}_v \left(2 \cdot \frac{{c_{uv}}/{2n}}{1 - c_{uv}/2n} (1 - (G_1)_{uv}) - \frac{c_{uv}}{n} \right) \label{eqn:jumbojet} \\
                              &= \sum \hat{x}_u \hat{x}_v \frac{c_{uv}^2/2n^2}{1- c_{uv}/2n} - \sum \hat{x}_u \hat{x}_v \frac{c_{uv}/n}{1- c_{uv}/2n} (G_1)_{uv} \\
    \implies |\mathbb{E}[T\mid G_1]| &\le \sum_{u<v}  \frac{c_{uv}^2/2n^2}{1- c_{uv}/2n} +  \sum_{u<v} \frac{c_{uv}/n}{1- c_{uv}/2n} (G_1)_{uv} \\
                                    &\le \frac{a^2/2n^2}{1 - a/2n} \binom{n}{2} + \frac{a/n}{1-a/2n} \#G_1.
    \end{align}
    
    where recall that $\#G_1$ is the number of edges in $G_1.$ Note that we may condition on $\mathcal{E},$ the occurrence of which is a deterministic function of $G_1.$ Since under $\mathcal{E}$ we have $\#G_1 \le an/2,$ we find that \begin{equation} \label{eqn:tst_null_centre}|\mathbb{E}[T\mid G_1, \mathcal{E}]| \le \frac{1}{1-a/2n} \left( \frac{a^2}{2n^2} \frac{n^2}{2}  +  \frac{a}{n} \frac{an}{2}\right) \le  a^2,\end{equation}
    where the final inequality uses that $1/(1-a/2n) \le 4/3,$ which follows from $a \le (n/2)^{1/3},$ and $n \ge 2.$
    
    Finally observe that the right hand side of the equation above does not depend on $G_1.$ Thus, we may integrate over $P(G_1\mid \mathcal{E})$ to find that \( |\mathbb{E}[T\mid \mathcal{E}]| \le a^2. \)
    
    \paragraph{Remark} This lemma is likely rather weak. In particular, the upper bound on $|\mathbb{E}[T|G_1]|$ completely ignores the relationship between $\hat{x}$ \& $G_1,$ and that between $G_1$ \& $c_{uv}$. Indeed, (\ref{eqn:jumbojet}) may also be rewritten as \[ \mathbb{E}[T\mid G_1] = \sum \frac{c_{uv}/n}{1-c_{uv}/2n} \hat{x}_u\hat{x}_v \left(\frac{c_{uv}}{2n} - (G_1)_{uv} \right).\] Since $(G_1)_{uv} \sim \textrm{Bern}(c_{uv}/2n),$ and $\hat{x}$ is a clustering derived from $G_1$, it may be possible to control the above to something much smaller than $a^2.$ This may require nontrivial use of the $\mathcal{E}(\hat{x})$ conditioning here, which is unused in the above argument. Unfortunately it seems that such control would closely depend on the scheme used to obtain $\hat{x},$ which tend to be complex - most schemes involve non-trivial regularisation of $G_1,$ as well as some amount of quantisation of the solution to an optimisation problem to produce $\hat{x},$ due to which the covariance of $G_1$ and $\hat{x}$ is difficult to understand. For completeness' sake we point out that an upper bound on the same of $O(a^2/n)$ would remove the nuisance condition of $a \le n^{1/3}$ present in Theorem \ref{thm:tst}. \end{proof}

\subsubsection{Setting up the Proof of Lemma \ref{lem:tst_integrated}} \label{appx:tst_main}

\begin{proof}[\hspace{-6pt}\nopunct]
\let\qed\relax

We proceed by first developing some intuition behind the proof of Lemma \ref{lem:tst_integrated} instead of launching straight into the same. Further, we assume throughout that $d(x,y) \ge s.$ 

Let \begin{align*}
    \textrm{Incorrect} &:= \{ u \in [1:n]: x(u) \neq \hat{x}(u) \} \\
    \textrm{Unchanged} &:= \{u \in [1:n]: x(u) = y(u) \}.
\end{align*}
  
  and the sets `\textrm{Correct}' and `\textrm{Changed}' be their respective complements. We show in Appendix \ref{appx: tst} the following lemma 
  \begin{mylem}\label{lem:tstmean}\emph{
        \begin{align}
    \mathbb{E}[T^{\hat{x}}(G') - T^{\hat{x}}(H) \mid \hat{x}]  = \frac{(a-b)}{n}  \Big{(} &n(\textrm{Unchanged}) - 2n(\textrm{Incorrect, Unchanged}) \Big{)}  \notag \\ &\times \Big{(} n(\textrm{Changed}) - 2n(\textrm{Incorrect, Changed}) \Big{)}, \end{align} } where \emph{ \begin{align*}
        n(\textrm{Unchanged}) &= |\textrm{Unchanged}|\\
        n(\textrm{Incorrect, Unchanged}) &= |\textrm{Incorrect} \cap \textrm{Unchanged}|,
    \end{align*}} and the other terms are defined analogously.
  \end{mylem}
   
   
   We note that the above lemma holds irrespective of the balance assumptions in the theorem. 
   
   Suppose $n(\textrm{Incorrect}) = k.$  Due to the exchangability of the nodes when $|\{u : x(u) = +\}| = |\{u: x(u) = -\}|,$ the incorrectly labelled nodes in $\hat{x}$ correspond to a choice of $k \in [0:n/2]$ nodes picked without replacement from $[1:n]$ uniformly at random. Further, since the changes made in $y$ are chosen independently of the graphs, they are independent of $\hat{x}.$ Thus, the number of correct and incorrect nodes changed forms hypergeometric distribution. The expected number of Incorrect nodes changed is precisely $\frac{s}{n} \cdot k,$ where $s$ is the number of changes made, and similarly for Incorrect nodes unchanged. 
   
   Further invoking the results of \cite{fei2019exponential}, if $ \Lambda \ge C \log(1/\varepsilon_{\max}),$ then $k \le \varepsilon_{max} n$ with probability at least $1 - 1/n.$ As a consequence, the bound in Lemma \ref{lem:tstmean} remains large in magnitude even on integrating over the randomness in $\hat{x}.$ This was the subject of Lemma \ref{lem:tst_integrated} from the text, reproduced below for convenience.\\
   
   \noindent \textbf{Lemma \ref{lem:tst_integrated}} 
    \begin{align*}
    \mathbb{E}[T^{\hat{x}}(G') - T^{\hat{x}}(H) \mid \mathcal{E}]  \ge  \left( (1-2\varepsilon_{\max})^2 - \frac{1}{n-1}\right) \frac{(a-b)}{n}( n -s )s, \end{align*}
   the proof of which is the subject of Appendix \ref{appx:tst_integrated}. 
\end{proof}

\subsubsection{Proof of Lemma \ref{lem:tstmean}}
\label{appx: tst}

\begin{proof}[\hspace{-6pt}\nopunct]
We will require explicit counting of a number of groups of nodes. Let us first define them:
  
  Let \begin{align*} S^{++} &:= \{ u\ \in [1:n] : \hat{x}(u) = +1, \, x(u) = +1 \}, \quad n^{++} := |S^{++}|, \\ 
                     S^{+-} &:= \{ u\ \in [1:n] : \hat{x}(u) = +1, \, x(u) = -1 \}, \quad n^{+-} := |S^{+-}|, \\
                     S^{--} &:= \{ u\ \in [1:n] : \hat{x}(u) = -1,\, x(u) = -1 \}, \quad n^{--} := |S^{--}|, \\ 
                     S^{-+} &:= \{ u\ \in [1:n] : \hat{x}(u) = -1,\, x(u) = +1 \}, \quad n^{-+} := |S^{-+}|.\end{align*} 
    
    The sets above encode the partitions induced by $\hat{x}$ and $x,$ with the first symbol in the superscript denoting the label given by $\hat{x}.$ Observe that $S^{+-}, S^{-+}$ are the sets of nodes mislabelled in $\hat{x}.$
    
    Lastly, for $(\mathfrak{i},\mathfrak{j} ) \in \{+, -\}^2,$ let \begin{align*}
        C^{\mathfrak{i},\mathfrak{j}} &:= S^{\mathfrak{i},\mathfrak{j}} \cap \{ u \in [1:n] : x(u) \neq y(u) \} \\
        n_C^{\mathfrak{i},\mathfrak{j}} &:= |C^{\mathfrak{i},\mathfrak{j}}|
    \end{align*} 
    
    These are the nodes that change their labels in $y.$ Note that the values of each of the above objects is a function of $\hat{x}.$ For now we will fix $\hat{x},$ and compute expectations over the randomness in $G', H$ alone.\\
    
    We first study $N_w$: $N_w^{\hat{x}}(G) = N_w^{\hat{x}}(G[+]) + N_w^{\hat{x}}(G[-]),$ where $G[+]$ is the induced subgraph on the nodes $\{u \in [1:n]: \hat{x}(u) = +\}$ and similarly $G[-].$ 
    
    By simple counting arguments, \begin{align} \mathbb{E}[N_w^{\hat{x}}(G'[+]) \mid \hat{x}] =  \binom{n^{++} + n^{+-}}{2} \frac{a}{n} - \frac{(a-b)}{n} n^{++} n^{+-}. \end{align}
   
    Under $H,$ the nodes in $C^{++}$ behave as if they were in $S^{+-}$ and those in $C^{+-}$ as if they were in $S^{++}$.  
     Computations analogous to before lead to \begin{equation} \mathbb{E}[N_w^{\hat{x}}(G'[+]) - N_w^{\hat{x}}(H[+])\mid \hat{x}] =  \frac{a-b}{n} \left((n^{++}-n_C^{++})  - (n^{+-} - n_C^{+-}) \right)(n_C^{++} - n_C^{+-})  \end{equation}
    
    By symmetry, we can obtain the above for $G[-]$s by toggling the group labels above. Thus, conditioned on a fixed $\hat{x},$ we have \begin{align} \mathbb{E}[N_w^{\hat{x}}(G') - N_w^{\hat{x}}(H) \mid \hat{x}] =  &\frac{(a-b)}{n} {\Big(} \left((n^{++}-n_C^{++})  - (n^{+-} - n_C^{+-})\right)(n_C^{++} - n_C^{+-}) \notag \\    &\quad\qquad +  \left( (n^{--}-n_C^{--})  - (n^{-+} - n_C^{-+})\right) (n_C^{--} - n_C^{-+}) {\Big )}.\label{eq:bigmess_w} \end{align}
    
    Similar calculations can be performed for $N_a$. Since in edges across the true partitions, the edges in the same group appear with probability $a/n$ and in different groups with $b/n,$ the roles of $a$ and $b$ will be exchanged in this case, leading to a factor of $+(a-b)$ instead of $-(a-b).$ We will suppress the tedious computations, and simply state that 
    \begin{align} \mathbb{E}[N_a^{\hat{x}}(G') - N_a^{\hat{x}}(H) \mid \hat{x}] =  &\frac{(a-b)}{n} {\Big(}\left((n^{++}-n_C^{++})  - (n^{+-} - n_C^{+-})\right) (n_C^{--} - n_C^{-+}) \notag\\    &\quad\qquad +  \left((n^{--}-n_C^{--})  - (n^{-+} - n_C^{-+})\right) (n_C^{++} - n_C^{+-}) {\Big)}. \label{eq:bigmess_a}\end{align}
        
    For convenience, we define \begin{alignat*}{2}
        &n(\textrm{Correct, Unchanged}) &&:= (n^{++} + n^{--})  - (n^{++}_C + n^{--}_C) \\
        &n(\textrm{Correct, Changed}) &&:= (n^{++}_C + n^{--}_C) \\
        &n(\textrm{Incorrect, Unchanged}) &&:= (n^{+-} + n^{-+})  - (n^{+-}_C + n^{-+}_C) \\
        &n(\textrm{Incorrect, Changed}) &&:= (n^{+-}_C + n^{-+}_C) \\
    \end{alignat*}
    
    where `correctness' corresponds to the nodes $u$ such that $\hat{x}(u) = x(u),$ while `unchangedness' to $u$ such that $x(u) = y(u).$
    
    Subtracting (\ref{eq:bigmess_a}) from (\ref{eq:bigmess_w}) then yields that for fixed $\hat{x}$ \begin{align}\label{eq:mess_appx}
    \mathbb{E}[T^{\hat{x}}(G') - T^{\hat{x}}(H) \mid \hat{x}] = \frac{(a-b)}{n} {\Big(}  n(&\textrm{Correct, Unchanged}) - n(\textrm{Incorrect, Unchanged}) \Big{)}  \notag \\ & \times \Big{(}n(\textrm{Correct, Changed}) - n(\textrm{Incorrect, Changed}) {\Big)}. \end{align} 
    
    The lemma now follows on observing that \[ n(\textrm{Unchanged}) = n(\textrm{Correct, Unchanged}) + n(\textrm{Incorrect, Unchanged}),\] and similarly $n(\textrm{Changed}).$ 
\end{proof}

\subsubsection{Proof of Lemma \ref{lem:tst_integrated}}\label{appx:tst_integrated}

\begin{proof}[\hspace{-6pt}\nopunct]
    
    Below we will simply assume that $d(x,y) = s.$ The proof is easily extended to $>s$. 
    
    Effectively, we are considering the following process: we have a bag of $n$ balls - corresponding to the nodes - of two colours (types), Changed and Unchanged,  and we are picking $k \le n/2$ of them uniformly at random without replacement. Let \begin{align}
        \eta_1 &:= n(\textrm{Unchanged, Incorrect})\\
        \eta_2 &:= n(\textrm{Changed, Incorrect})
    \end{align}
    
    and 
    \begin{align}  \zeta :=& (n(\textrm{Unchanged}) - 2n(\textrm{Incorrect, Unchanged}))(n(\textrm{Changed}) - 2n(\textrm{Incorrect, Changed}))  \notag \\
                          =& (n - s - 2\eta_1)(s - 2\eta_2).\end{align}
    
    We now condition on the number of errors being $k$, which imposes the condition that $\eta_1 + \eta_2 = k$. Recall the sampling without replacement distribution, which implies that \begin{align}
        P(\eta_1 = k-j, \eta_2 = j \mid d(\hat{x}, x) = k) = \frac{\binom{n-s}{k-j} \binom{s}{j}}{\binom{n}{k}}.
    \end{align}                      
    
    Thus, \begin{align*}
        \mathbb{E}[\eta_1 | d(x, \hat{x}) = k] &= \frac{k}{n} (n-s) \\
        \mathbb{E}[\eta_2 | d(x, \hat{x}) = k] &= \frac{k}{n}(s) \\
        \mathbb{E}[\eta_1 \eta_2 | d(x, \hat{x}) = k] &= (n-s)(s) \frac{k(k-1)}{n(n-1)} = s(n-s) \left( \frac{k^2}{n^2} - \frac{k(n-k)}{n^2(n-1)}\right).
    \end{align*}
    
    As a consequence, we obtain that \begin{align}
        \mathbb{E}[\zeta | d(x, \hat{x}) = k] &= s(n-s) \left( 1 - 4\frac{k}{n} + 4\frac{k^2}{n^2} - 4\frac{k(n-k)}{n^2(n-1)}\right) \notag\\
                                              &= s(n-s) \left( \left(1 - 2\frac{k}{n}\right)^2 - 4\frac{k(n-k)}{n^2(n-1)} \right)      
    \end{align}
    
    Note that the above is decreasing as $k$ increases for $k \le n/2.$ 
    
    Note further that the Markov chain $\zeta\text{--}d(\hat{x},x)\text{--}G_1$ holds. Thus the above also holds for $\mathbb{E}[\zeta \mid \mathcal{E}(G_1),d(x, \hat{x}) = k ].$
    
    We now condition on $\mathcal{E}(\hat{x})$ to find that 
    \begin{align}
        \frac{\mathbb{E}[\zeta\mid \mathcal{E}(\hat{x}), \mathcal{E}(G_1)]}{s(n-s)} &\ge \left( (1 - 2\varepsilon_{\max})^2 - 4\frac{\varepsilon_{\max}(1- \varepsilon_{\max})}{n-1} \right) \\
                                         &\ge (1- 2\varepsilon_{\max})^2 - \frac{1}{n-1}
    \end{align}
    
    where we have used $\varepsilon_{\max} \le 1/2,$ and the (unstated but obvious) condition that $n \ge 2.$ 
    
    Applying the above to the result of Lemma \ref{lem:tstmean}, we find that     
    \[
        \mathbb{E}[T^{\hat{x}}(G') - T^{\hat{x}}(H) \mid \mathcal{E}]  \ge \left( (1-2\varepsilon_{\max})^2 - \frac{1}{n-1}\right) \frac{(a-b)}{n}( n -s )s.  \qedhere
    \]     
\end{proof}

\subsubsection{Proof of Lemma \ref{lem:tst_unbal_alt}}\label{appx:tst_ubal_pf}

\begin{proof}[\hspace{-6pt}\nopunct] For continuity of exposition, we prove Lemma \ref{lem:tst_unbal_alt} before Lemma \ref{tst:conc}.

Below we will simply assume that $d(x,y) = s.$ The proof is easily extended to the same being $>s$.

We begin with recalling Lemma \ref{lem:tstmean}, and noting that it's proof does not utilise the exact balance assumption. We begin as in the proof of Lemma \ref{lem:tst_integrated}, by defining  \begin{align*}
        \eta_1 &:= n(\textrm{Unchanged, Incorrect})\\
        \eta_2 &:= n(\textrm{Changed, Incorrect})
    \end{align*}
    
    and noting from Lemma \ref{lem:tstmean} that $\mathbb{E}[T^{\hat{x}}(G') - T^{\hat{x}}(H)\mid \hat{x}] \ge \frac{(a-b)}{n} (n-s - 2\eta_1) (s-2\eta_2) =: \frac{(a-b)}{n} \zeta.$ Once again, let's fix the errors number of errors made by $\hat{x}$ as some $k$, and let $s_+, s_-$ be the number of changes made in communities $+$ and $-$ respectively. Note that $s_+ + s_- = s.$
    
    Using the above definitions, \[ \zeta = (n-s)(s - 2\eta_2) - 2s \eta_1 + 4\eta_1\eta_2 \ge \frac{n}{2} (s-2\eta_2) - 2ks, \] by noting that $\eta_1\eta_2\ge 0, \eta_1 \le k$ and that $s \le n/2.$ Thus, \[ \mathbb{E}[\zeta| d(\hat{x}, x) \le k] \ge \frac{ns}{2} \left(1 - \frac{4k}{n} - 2\frac{\mathbb{E}[\eta_2|d(\hat{x},x) \le k]}{s} \right) .\]
    
    Suppose that the recovery procedure makes $(k^+, k^-)$-errors in communities $+$ and $-$ respectively, with $k^+ + k^- \le k$.  \emph{Within-community} exchangability of nodes implies that the errors made within a community must be uniformly distributed over the community. Since the changes are made independently of these errors, we must have that the number of changed nodes in community $* \in \{+1, -1\}$ that are incorrectly inferred must be $\mathrm{Hyp}(n_*, s_*, k^*)$ distributed. In particular, this yields that \[\mathbb{E}[\eta_2| (k^+, k^-)] =   \frac{s_+}{n_+} k^+ + \frac{s_-}{n_-} k^- \le \frac{s_+ k ^+ + s_- k^-}{\min(n_+, n_-)} \le \frac{k^+ + k^-}{\min(n_+, n_-)} s. \]
    
    The above immediately yields that \[ \mathbb{E}[\zeta| d(\hat{x}, x) \le k] \ge \frac{ns}{2} \left(1 - \frac{4k}{n} - 2\frac{k}{\min(n_+, n_-)} \right) .\] On $\mathcal{E}, k \le \varepsilon n,$ leading to the claimed bound. \end{proof}

\subsubsection{Proof of Lemma \ref{tst:conc} }\label{appx:tst_conc}\begin{proof}[\hspace{-6pt}\nopunct]

Recall the notation from Appendix \ref{appx:tst_null_centre}. Under the null $H \overset{\textrm{law}}{=} G'.$ Below, we will use $G'$ as a proxy for $H$ in the null distribution, and use $H$ only in the alternate.

To begin with, observe that both $G', H$ are independent of $G_1, \widetilde{G}, \hat{x},$ and that $\widetilde{G}$ is independent of $\hat{x}$ given $G_1.$ Now, $T^{\hat{x}}$ is a signed sum of independent Bernoulli random variables with parameters smaller than $a/n$ given $G_1.$. Thus, invoking results from Ch. 2 of \cite{Chung:2006:CGN:1208774} (and using that for $a \ge C$ for some large enough $C$ implies that $a \ge 16 \log(6n)/n \iff 1/6n \le \exp{-na/16}$)), we find that for $\Gamma \in \{\widetilde{G}, G', H\},$ \[ P\left( \left|T^{\hat{x}}(\Gamma) - \mathbb{E}[T^{\hat{x}}(\Gamma)\mid G_1, \mathcal{E}] \right| \ge \sqrt{2na \log(6n)} \mid G_1, \mathcal{E}\right) \le \frac{1}{3n},\] where we have used that $\mathcal{E}$ is determined given $G_1$ (i.e. $\mathcal{E}$ lies in the sigma-algebra generated by $G_1.$) 

We now control the null and alternate fluctuations given $\mathcal{E}.$
\begin{enumerate}
    \item[Null:]  By the union bound, we find that \[ P\left( 2T^{\hat{x}} (\widetilde{G}) - T^{\hat{x}}(G') \ge \mathbb{E}[2T^{\hat{x}} (\widetilde{G}) - T^{\hat{x}}(G') \mid G_1, \mathcal{E}] + 3\sqrt{2na \log(6n)}  \mid G_1, \mathcal{E}\right) \le \frac{2}{3n} \]
    
    Recall from equation (\ref{eqn:tst_null_centre}) from the proof of Lemma \ref{lem:tst_null_centre} that $ \mathbb{E}[2T^{\hat{x}} (\widetilde{G}) - T^{\hat{x}}(G') \mid G_1, \mathcal{E}] \le a^2.$ Feeding this in, we find that 
    \[ P\left( 2T^{\hat{x}} (\widetilde{G}) - T^{\hat{x}}(G') \ge a^2 + 3\sqrt{2na \log(6n)}  \mid G_1, \mathcal{E}\right) \le \frac{2}{3n}. \]
    The right hand side above does not depend on $G_1,$ and neither does the fluctuation radius wihtin the probability. Thus integrating over $P(G_1\mid \mathcal{E}),$ we find that \[ P\left( T \ge a^2 + 3\sqrt{2na \log(6n)}  \mid \mathcal{E}\right) \le \frac{2}{3n}, \] where we have used that $T = 2T^{\hat{x}} (\widetilde{G}) - T^{\hat{x}}(H) \overset{\mathrm{law}}{=} 2T^{\hat{x}} (\widetilde{G}) - T^{\hat{x}}(G')$ under the null.
    
    \item[Alt:] Following the above development again, this time with lower tails, we find that given $G_1$ with probability at least $1 - 2/3n,$ \begin{align*} 2T^{\hat{x}} (\widetilde{G}) - T^{\hat{x}}(G') &\ge -(\mathbb{E}[2T^{\hat{x}} (\widetilde{G}) - T^{\hat{x}}(G') \mid G_1, \mathcal{E}]) - 3\sqrt{2na \log(6n)} \\ T^{\hat{x}}(G') - T^{\hat{x}}(H) &\ge +(\mathbb{E}[T^{\hat{x}}(G') - T^{\hat{x}}(H) \mid G_1, \mathcal{E}]) - 2\sqrt{2na \log(6n)} \end{align*}
    
    Further, given $(G_1,\mathcal{E}),$ by Lemmas \ref{lem:tst_null_centre}, \ref{lem:tst_integrated} we have \begin{align*} 2T^{\hat{x}} (\widetilde{G}) - T^{\hat{x}}(G') &\ge -a^2 - 3\sqrt{2na \log(6n)} \\ T^{\hat{x}}(G') - T^{\hat{x}}(H) &\ge +\kappa (a-b) s (1-s/n)  - 2\sqrt{2na \log(6n)}, \end{align*}
    
    where $\kappa = (1 - 2\varepsilon_{\max})^2 - 1/(n-1).$ Adding the above, we find by the union bound that \[ P\bigg(2T^{\hat{x}} (\widetilde{G}) - T^{\hat{x}}(H) \ge \kappa (a-b) s(1-s/n) -a^2 - 5\sqrt{2na \log(6n)} \mid G_1, \mathcal{E} \bigg) \ge 1 - \frac{4}{3n}.\] The claim follows on noting that the right hand side and the fluctuation radius do not depend on $G_1,$ and integrating the inequality over $G_1$. \qedhere
\end{enumerate}

\end{proof}

\subsection{Proof of the converse bound from Theorem \ref{thm:tst}.}\label{appx:tst_lower}

We restate the lower bound below as a proposition:

\begin{myprop*}
    There exists a universal constant $C $, and another $c < 1$ that depends on $C$, such that if $\Lambda \le C$ and $s \le  \frac{n}{2}(1-c),$ then reliable two-sample testing of balanced communities for $s$ changes is impossible for large enough $n$. 
    
    In particular, for $a+b < n/4,$  the statement holds with $C = 1/8, c = 1/6$ for $n \ge 136,$ and in this case, $R_{\mathrm{TST}} \ge 0.25$.    
\end{myprop*}

\begin{proof}
The proof proceeds by using a variation of Le Cam's method, and importing impossibility results for the so-called distinguishability problem \cite{banks2016information}. In particular, suppose that in the null distribution, the communities are drawn according to the uniform prior on balanced communities, denoted by $\pi.$ Further, assume that if a $s$-change is made, then the resulting community is chosen uniformly from all communities that are at least $s$ far from the null community. We have the hypothesis test: \begin{align*}
    &H_0: (G,H) \sim \sum_{x \in \mathcal{B}} \pi_x P(G|x) P(H|x) &\textrm{vs } &H_1: (G,H) \sim \sum_{x,y \in \mathcal{B}} \pi_x \pi_{y|x} P(G|x) P(H|y),
\end{align*}

where we use $\mathcal{B}$ to denote the set of balanced communities, and $\pi_{y|x} $ is the uniform distribution on $\mathcal{B} \cap \{y: d(x,y) \ge s\}.$ For succinctness, let us denote the null and alternate distributions above as $p_{\mathrm{null}}$ and $p_{\mathrm{alt}}$ respectively.

Once again, by Neyman-Pearson theory, \[ R_{\mathrm{TST}} \ge R_\pi \ge  1 - d_{\mathrm{TV}}(p_{\mathrm{null}}, p_{\mathrm{alt}}) \ge 1 - d_{\mathrm{TV}}(p_{\mathrm{null}}, Q) - d_{\mathrm{TV}}(Q, p_{\mathrm{alt}}),  \] where $Q$ is any distribution, and the last inequality is since $d_{\mathrm{TV}}$ is metric. 

We choose $Q$ to be the unstructured distribution induced by an Erd\H{o}s-R\'{e}nyi graph of parameter $(a+b)/2n.$ The primary reason for this is that explicit control on the total variation distance between $p_{\mathrm{null}}$ and $Q$ is then available - for instance, by \cite[\S3.1.2]{wu2018statistical}, we have \[ D_{\chi^2}(p_{\mathrm{null} } \|Q) + 1 \le \mathbb{E}\left[ \exp{ \tau \left( \frac{4 \mathscr{H} - n}{\sqrt{n}} \right)^2} \right], \] where $\mathscr{H}$ is a $ \mathrm{Hypergeometric}(n, n/2, n/2)$ random variable, and \[ \tau =  \frac{(a-b)^2}{2(a+b)} + \frac{(a-b)^2}{2(2 - a/n - b/n)} .\] Notice the extra factor of $2$ compared to the expressions in \cite{wu2018statistical}, which arises since we sum over two independent graphs $G,H$ and not one. We observe that \[ \tau = \Lambda \frac{n}{2n - a - b} ,\] and explicitly, if $ a+b \le n/4,$ then $\tau \le \frac{4}{7} \Lambda$. 

We now consider the alternate term. As a preliminary, let \[ \gamma:= \frac{ \sum_{k = 0}^{s-1} \binom{n/2}{k/2}^2}{ \binom{n}{n/2}}.  \] Note that $\gamma$ is the probability that two balanced communities chosen independently and uniformly, lie within distortion $s$. Indeed, since communities are formed by identifying antipodal points in the boolean cube, the probability of picking a community at distortion $< s$ coincides with that of picking a balanced vector at Hamming distance $< s$ from a given balanced vector in the cube $\{0,1\}^n$. The denominator in $\gamma$ is clearly the number of balanced vectors in the cube, while the numerator is the number of balanced vectors at a distance of $< s$ from any given balanced vector - we choose $k < s,$ and choose $k/2$ points marked $1$ and $k/2$ marked $0$, and flip them all. 

As a consequence, we find that for any $x,y \in \mathcal{B},$ \[ \pi_{y|x} \le \frac{\pi_y}{1 - \gamma} .\]    

Thus, in the $\chi^2$ expressions for $p_{\mathrm{alt}},$ we have \begin{align*}
    \mathbb{E}_{(G,H) \sim Q^{\otimes 2}} [ (p_{\mathrm{alt}}/Q)^2] &= \sum_{x,y,x', y'} \mathbb{E}\left[\frac{P(G|x)P(G|x')}{Q^2(G)}\frac{P(H|y)P(H|y')}{Q^2(H)}\right] \pi_x \pi_{x'} \pi_{y|x} \pi_{y'|x'} \\ 
                                                                    &\le \frac{1}{(1-\gamma)^2} \sum_{x,y,x', y'} \mathbb{E}\left[\frac{P(G|x)P(G|x')}{Q^2(G)}\frac{P(H|y)P(H|y')}{Q^2(H)}\right] \pi_x \pi_{x'} \pi_{y} \pi_{y'} \\
                                                                    &= \frac{1}{(1-\gamma)^2} \left(1 + \chi^2( \sum_{x \in \mathcal{B}} \pi_x P(G|x) \| Q (G) ) \right)^2
\end{align*}

Since the final quantity is explicitly controlled in the cited section, we also have \[ 1 + D_{\chi^2}(p_{\mathrm{alt}} \| Q) \le \frac{1}{(1-\gamma)^2} \mathbb{E}\left[  \exp{ \frac{\tau}{2} \left( \frac{4\mathscr{H} - n}{\sqrt{n}} \right)^2 }  \right]^2 \le \frac{1}{(1-\gamma)^2} \mathbb{E}\left[  \exp{ {\tau}\left( \frac{4\mathscr{H} - n}{\sqrt{n}} \right)^2 }  \right], \]

the final relation arising from Jensen's inequality.

Since the quantity appears often, we let \[ \beta:= \mathbb{E}\left[  \exp{ {\tau}\left( \frac{4\mathscr{H} - n}{\sqrt{n}} \right)^2 }  \right].\] Invoking the inequality $d_{\mathrm{TV}} \le \sqrt{\log(1 + D_{\chi^2})/2},$ we find that \[ R_{\mathrm{TST}} \ge 1 - \sqrt{\log(\beta)/2} - \sqrt{\log(\beta(1-\gamma)^{-2})/2} = 1 - \sqrt{\log(\beta/(1-\gamma) )}.  \]

Note that the only $s$-dependent term in the above bounds is $\gamma.$ We first offer control on the $\gamma$, and claim that for $s/n < 1/2$, $\gamma \to 0$.  Indeed, since $s \le n/2$, and by standard refinements of Stirling's approximation (for instance, we use \cite[Exercise 5.8]{gallager1968information} below), \begin{align*}  \gamma &\le s \frac{ \binom{n/2}{s/2}^2}{\binom{n}{n/2}} \le s \frac{1}{2\pi} \frac{n/2}{s/2 (n-s)/2} 2^{n h_2(s/n) }  \left( \sqrt{\frac{n}{8 (n/2)^2}} 2^n \right)^{-1} \\ &\le \sqrt{\frac{2n}{\pi^2} } 2^{- n( 1- h_2(s/n) )},\end{align*}

where $h_2$ is binary entropy in bits. 

At this point the argument in the limit as $n \to \infty$ is complete - since $4(\mathscr{H} - n)/\sqrt{2n}  \overset{\mathrm{Law}}{\to} \mathcal{N}(0,1)$, $\beta$ is bounded as $n \to \infty$ by $\sqrt{1 - 2\tau}$ if $\tau < 1/2$, and since in this limit $\tau \to \Lambda/2$ (for $a,b = o(n)$), we obtain that if $\limsup s/n < 1/2,$ and $\Lambda < 1,$ then $\liminf R_{\mathrm{TST}} > 0.$

Non-asymptotic bounds can be recovered by giving up space on the constants, leading to the statement we have claimed. 

Concretely, to attain $R_{\mathrm{TST}} > 1/4,$ it suffice to show that $\beta (1-\gamma)^{-1} < e^{9/16}.$ Now, for $s < n/3,$ we have \[  (1 - \gamma) e^{9/16} \ge  \left( 1 - \sqrt{2n/\pi^2} 2^{-0.08 n} \right) e^{9/16} > 1.75\] for $n \ge 136.$\footnote{This is calculated using a computer algebra system. Analytically it is still easy to argue something similar - for $n > 10,$ $\sqrt{n}2^{-0.08 n}$ is decreasing. Thus, for $n > 100,$ $\sqrt{2n/\pi^2} 2^{-0.08 n} < \nicefrac{10\sqrt{2}}{256\pi} < 1/50, $ and thus the expression is at least $e^{9/16}\cdot\nicefrac{49}{50} > 1.6 \cdot \nicefrac{49}{50} > 1.5$ By following the next footnote with this number, this leads to an analytic proof of the conclusion holding for $\Lambda < \nicefrac{7}{80} \approx 0.087$. } Thus, it suffices to control $\beta$ to below $1.75$ in this regime. To this end, note that $u \mapsto \exp{ \tau ( (4u - n)/\sqrt{n})^2 }$ is a continuous, convex map, and thus, by \cite[Thm. 4]{Hoeffding}, \[ \beta \le \mathbb{E}\left[  \exp{ {\tau}\left( \frac{4\mathscr{B} - n}{\sqrt{n}} \right)^2 } \right], \] where $\mathscr{B} \sim \mathrm{Bin}(n/2, 1/2)$.

By Chernoff's bound, $P( |\mathscr{B} - n/4| \ge \sqrt{n}u) \le 2e^{-4u^2},$ and thus, we have \begin{align*}
    \beta &\le \int_{0}^{\infty} P\left( \exp{ {\tau}\left( \frac{4\mathscr{B} - n}{\sqrt{n}} \right)^2} \ge u \right) \,\mathrm{d}u \\
          &\le \int_0^\infty \min(1, 2u^{-1/4\tau}) \,\mathrm{d}\tau \\
          &= \frac{2^{4\tau}}{1- 4\tau} ,
\end{align*}

the final equality holding so long as $1/4\tau > 1 \iff \tau < 1/4.$ The original claim follows if \[ \frac{2^{4\tau}}{1- 4\tau} \le \frac{7}{4},\] which is true for $\tau < 0.074$. Since $\tau \le \nicefrac{4}{7} \Lambda, \Lambda < 1/8$ implies that $\tau < 4/56 < 0.072.$\footnote{The number $0.074$ is calculated using a computer algebra system. Purely analytic calculations are straightforward as well - for example by using $2^{4\tau} \le 1 + 4\tau$ for $\tau < 1/4,$ which can be proved by noting that $1+4\tau - 2^{4\tau}$ is initially increasing, and then strictly decreasing after a point, and that $1/4$ is a root of this function. This implies that the conclusion holds so long as $\tau < 3/44,$ which holds if $\Lambda < 21/176 \approx 0.119$.} 
\end{proof}

A couple of quick comments are useful here:\begin{enumerate}
    \item Note that the above cannot be applied usefully to GoF. This is because in GoF, the null is explicitly available, and we do not have the benefit of averaging with $\pi$ in the TV expressions. This causes the equivalent term $\chi^2(P(G|x_0) \|Q(G))$ to grow exponentially with $n\Lambda$. 
    \item The above characterises the tightness of our claimed bounds for TST of large changes - the method works if $\Lambda = \Omega(1)$ and $s \gg \sqrt{n \log n},$ and by the above argument, no test can work if $\Lambda \ll 1,$ as long as the change is not extreme ($ \limsup s/n < 1/2$ ). 
    \item While the above approach is wasteful in how it utilises $s$, this is actually a non-issue, since the bounds require a separate control on $d_{\mathrm{TV}}(p_{\mathrm{null}} \|Q),$ which can only be controlled if $\Lambda = O(1).$ In particular, we cannot pull out better bounds for the small $s$ situation from the above. 
\end{enumerate}

\section{Experimental Details}\label{appx:exp}

\subsection{Experiments on SBMs}\label{appx:exp_sbm}

The experiemnts simulate an ensemble of GoF and TST test and evaluate the performance of the two schemes using the sum of false alarm and missed detection probabilities ($FA + MD$). 

While the GoF scheme is implemented precisely as in the main text, the experiments use a slightly modified version of Algorithm~\ref{alg:tst} for the TST:
\begin{enumerate}[label=(\roman*)]
	\item $G_1$ subsamples $G$ at a rate $\eta$, and the test statistic $T$ is appropriately modified: $T \coloneqq \frac{1}{1-\eta} T^{\hat{x}_1}(\widetilde{G}) - T^{\hat{x}_1}(H)$. Intrinsically, the spectral clustering step is the more singal-sensitive part of the scheme \ref{alg:tst}. While splitting the graphs equally is fine for theoretical results, it is better in practice to devote more SNR to the clustering step, and less to compute the test statistic, which can be done by increasing $\eta$. In the following, we set $\eta = 0.85$. Other values of $\eta$ are explored in Appendix~\ref{appx:ext-expts-eta}.
	\item The constant factor in the threshold developed in the test is conservative, and we vary it to adjust for different values of $\eta$ and to mitigate its suboptimality. In the experiments, we used the threshold $\frac{3}{4}\sqrt{n (a+b) \log(6n)}$.
\end{enumerate}

As noted in the main text, the experiments are performed for various $(s, \Lambda)$ for a fixed value of $a/b = 3.$ $\Lambda$ is varied between $\Lambda_0$ and $10\Lambda_0$ for $\Lambda_0 = 3/4 log(n/100) \approx 1.7$. This is significantly below the theoretical threshold of $2$ necessary for non-trivial recovery. Further, $8\Lambda_0 = 2\log(n),$ at which point recovery with constant order distortion becomes viable.

\subsubsection{Implementation details} \label{appx:sbm-expt-details}

The experiment is setup as follows:
\begin{enumerate}
	\item We fix a value of $\Lambda_0 = 3/4 \log(n/100)$ as above. Then, for some choice of $b / a = r$, we choose $(a, b)$ satisfying $\Lambda = \alpha\Lambda_0$ and $\alpha \in [1, 10]$. $r$ is set to be $1/3$.
	\item For a fixed number of nodes, $n$, and for $s \in [1:n/2]$, we consider the balanced partition $x = [x_i]_{i=1}^n$ with
	\begin{align*}
	    x_i &= \begin{cases}
	        0, \quad 0 \leq i \leq n/2 \\
	        1, \quad n/2 < i \leq n
	    \end{cases}
	\intertext{for the null distribution, and the shifted balanced partition $y = [y_i]_{i=1}^n$ with}
	    y_i &= \begin{cases}
	        0, \quad s/2 < i \leq n/2+s/2 \\
	        1, \quad i \in (n/2, n] \cup [0, s/2]
	    \end{cases}
	\end{align*}
	for the alternate distribution. This ensures that $d(x, y) = s$. We take $\lfloor\cdot\rfloor$ whenever $s$ or $n$ are odd.
	\item We sample $G, G' \sim P(\cdot\,\vert\,x)$ and $H \sim P(\cdot\,\vert\,y)$, where $P$ represents drawing from an SBM with parameters $n$, $a$ and $b$, as described in \S\ref{sec:defi}
\end{enumerate}

\paragraph{GoF procedure.} Recall that we are given a proposed partition $x_0$. Here we set $x_0 = x$. The results of running the tests on the graph $G$ then serve to characterise size, and those on $H$ serve to characterise power.
\begin{enumerate}
    \item For the na\"{i}ve scheme, we produce partitions $\hat{x}$ and $\hat{y}$ from $G$ and $H$ respectively via spectral clustering (see below for details), and declare for null in either case if $d(x_0, z) < s / 2$, where $z$ is respectively $\hat{x}$ and $\hat{y}$.
    \item For the alternate scheme, we instead compute the statistic from \S\ref{sec:gof}, and reject on the basis of the threshold developed there.
\end{enumerate}

\paragraph{TST procedure.} Similarly to the above, runs on the pair $(G,G')$ serve to characterise size, and on $(G, H)$ serve to characterise power of the test. Precisely: 

\begin{enumerate}
    \item For the na\"ive two-sample test based on recovery and comparison, we estimate $\hat x, \hat x'$ and $\hat y$ from $G$, $G'$ and $H$ respectively. The structure is estimated using spectral clustering (see below for implementation details). We declare that a change has occurred if $d(\hat x, \hat x') \geq s / 2$, and no change if $d(\hat x, \hat y) < s / 2$. We get a false alarm every time we declare a change on the pair $(G, G')$, and we miss a detection whenever we declare no change on the pair $(G, H)$. The false alarm and missed detection probabilities are estimated as an average over $M = 100$ samples.
	\item For the two-sample test based on Algorithm~\ref{alg:tst}, we follow the algorithm as stated, making only the modifications previously described. To be precise, we estimate $\hat x_1$ from $G_1$, a subsampling of (the edges of) $G$ at rate $\eta$. Then, we compute the test statistics in the null and alternate distributions:
	\begin{align*}
	    T_{\text{Null}} &= \frac{1}{1-\eta}T^{\hat x_1}(\widetilde G) - T^{\hat x_1}(G')
	    \shortintertext{and}
	    T_{\text{Alt.}} &= \frac{1}{1-\eta}T^{\hat x_1}(\widetilde G) - T^{\hat x_1}(H),
	\end{align*}
	where $\tilde G = G - G_1$.
\end{enumerate}

In both the above cases, the simulations are performed over a range of $\Lambda = \alpha \Lambda_0$ and $s$, where $\alpha \in [1, 10]$ and $s \in (0, 250)$. Performance is indicated using the sum of false alarm and missed detection rates.

Details associated with the implementation of the aforementioned schemes are given below:
\begin{enumerate}
    \item All experiments were implemented in the Python programming language (v3.5+), using the Numpy (v1.12+) and Scipy (v0.18+) scientific computing packages~\cite{oliphant2006numpy,jones2001scipy}.
    \item Structure learning was performed using the Spectral Clustering~\cite{vonLuxburg2007tutorial} algorithm, as implemented by the Scikit-learn package (v0.19.1+)~\cite{pedregosa2011sklearn}.
    \item Spectral Clustering was regularized in the manner suggested by~\cite{joseph2016impact}. Effectively, if $G$ was the adjacency matrix to be submitted to the Scikit-learn spectral clustering function, we performed pre-addition, and instead passed $G + \tau \mathbf 1 \mathbf 1^\tpose$. We set $\tau = \frac{1}{10n}$, which proved sufficient to run the spectral clustering function with no errors or warnings.
    \item All plots were generated using Matplotlib~(v2.1+)~\cite{hunter2007matplotlib}.
\end{enumerate}

\subsubsection[Modifications to subsampling rate]{Modifications to $\eta$}
\label{appx:ext-expts-eta}

\begin{figure}[htb]
    \centering
    \begin{subfigure}[t]{0.48\textwidth}
        \includegraphics[width=\textwidth]{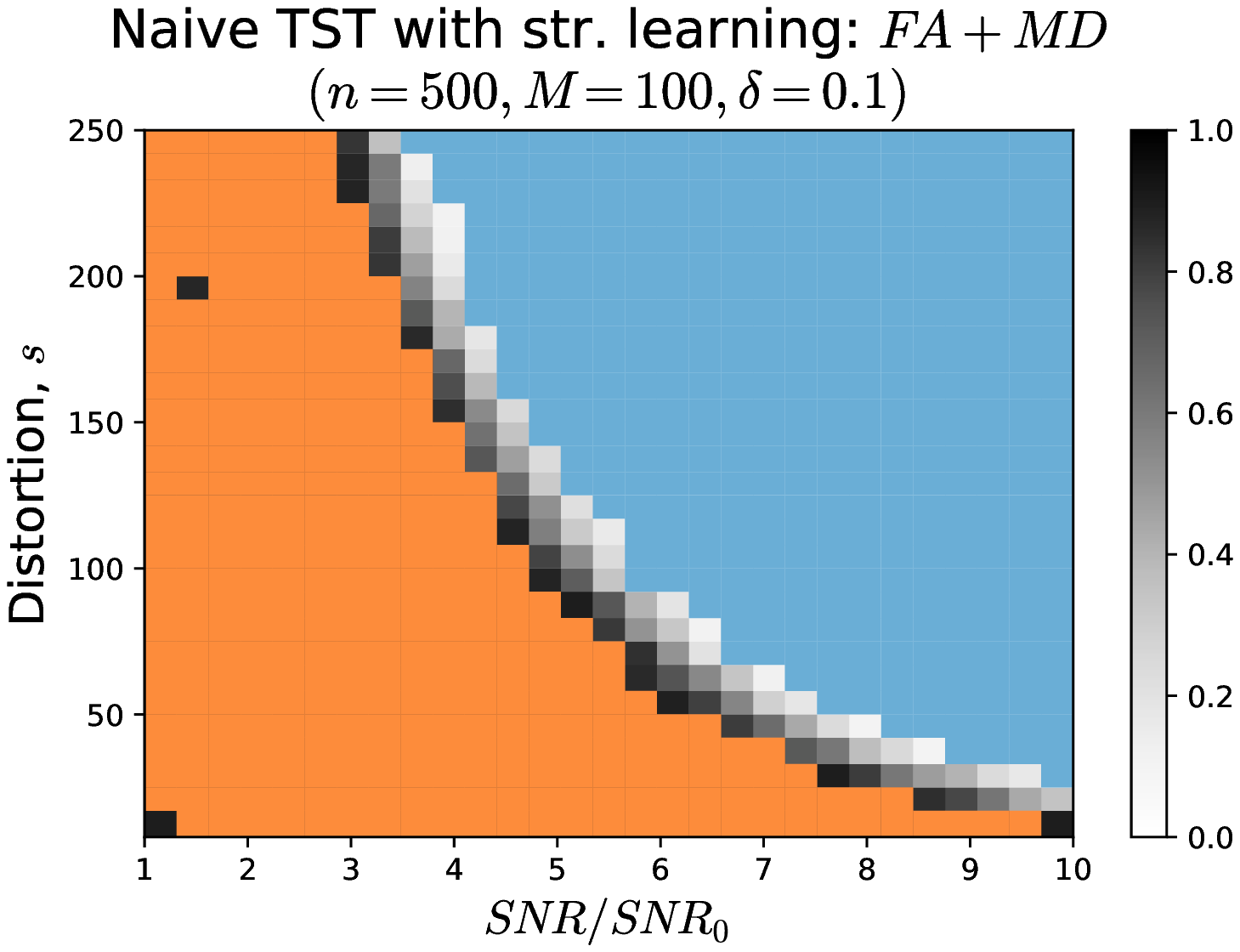}%
        \caption{Na\"ive two-sample test based on structure learning}
        \label{fig:sl-tst-eta}
    \end{subfigure}
    ~
    \begin{subfigure}[t]{0.48\textwidth}
        \includegraphics[width=\textwidth]{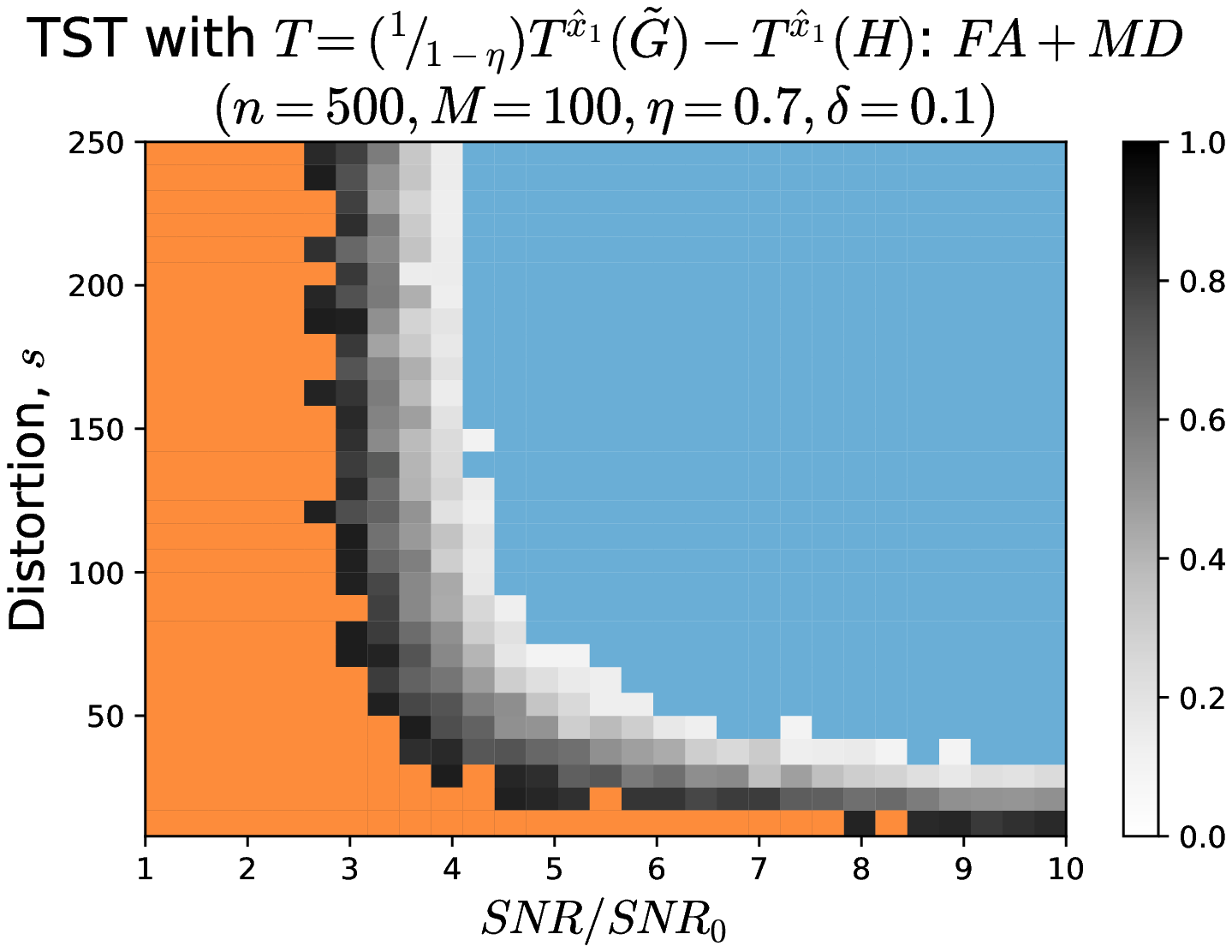}%
        \caption{Two-sample test based on Algorithm~\ref{alg:tst} for $\eta = 0.7$}
        \label{fig:tst-eta07}
    \end{subfigure}
    
    \begin{subfigure}[ht]{0.48\textwidth}
        \includegraphics[width=\textwidth]{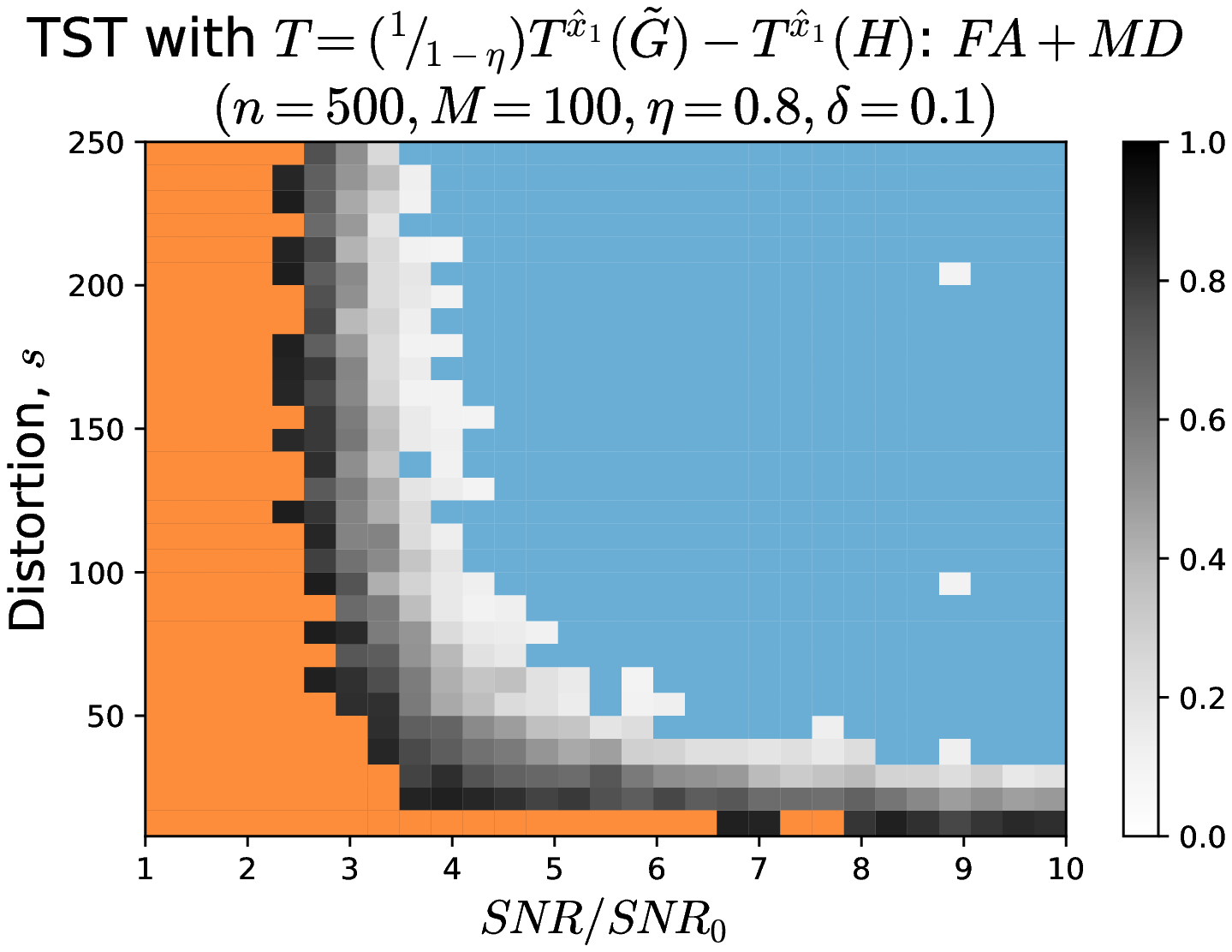}%
        \caption{Two-sample test based on Algorithm~\ref{alg:tst} for $\eta = 0.8$}
        \label{fig:tst-eta08}
    \end{subfigure}
    ~
    \begin{subfigure}[ht]{0.48\textwidth}
        \includegraphics[width=\textwidth]{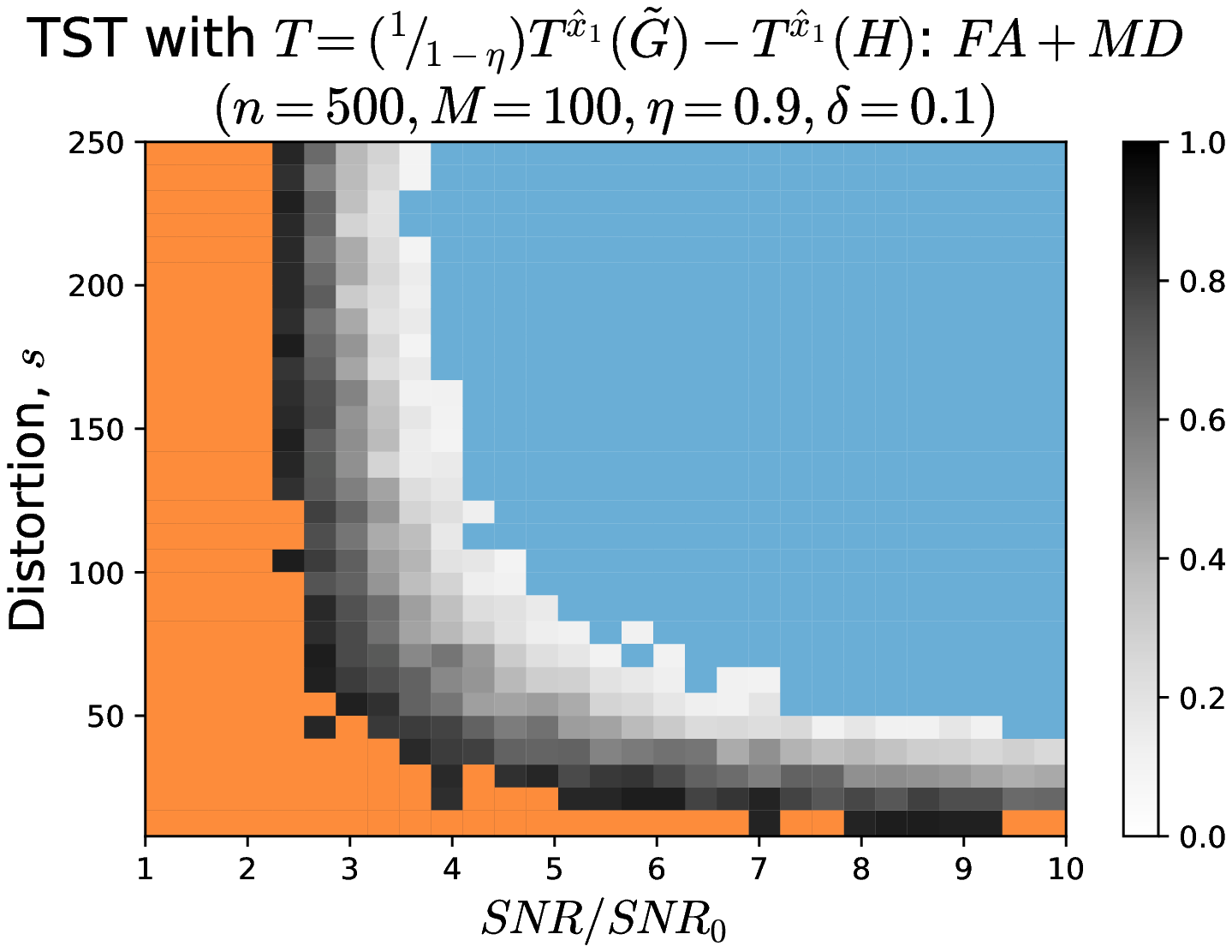}%
        \caption{Two-sample test based on Algorithm~\ref{alg:tst} for $\eta = 0.9$}
        \label{fig:tst-eta09}
    \end{subfigure}
    \caption{A comparison between the na\"ive two-sample test based on structure learning, and the two-sample test we propose in Algorithm~\ref{alg:tst}, for $\eta \in \{0.7, 0.8, 0.9\}$. Error rates lower than $\delta=0.1$ have been shaded blue to represent ``success'', while those higher than $1-\delta=0.9$ have been shaded orange to represent ``failure''.}
    \label{fig:sbm-eta}
\end{figure}

For completeness, we demonstrate how the performance of the modified two-sample test based on Algorithm~\ref{alg:tst} varies as $\eta$ is changed. Figure~\ref{fig:sbm-eta} compares the na\"ive two-sample test against the scheme based on Algorithm~\ref{alg:tst}, for three different values of $\eta$: $0.7$, $0.8$ and $0.9$.

We use the following parameters: $n = 500$, $\SNR_0 = \frac{3}{8}\log(n / 100) = \frac{3}{8}\log 5 \approx 0.5$, $\frac{b}{a} = r = 1/3$. For $\eta = 0.7$ and $\eta = 0.8$, the threshold used is $\sqrt{n (a+b) \log(6n)}$, while for $\eta = 0.9$, we used a higher threshold of $\frac{3}{2}\sqrt{n (a+b) \log(6n)}$.

While differences are rather subtle, a careful examination may reveal that as $\eta$ increases, the failure region recedes, while the success region advances in the high-$s$, low-$\SNR$ regime. However, the cost of this is an increased threshold to maintain success at $\delta=0.1$, and a wider transition region, indicating that different $\eta$ might be optimal at different $n$.

\subsection{Experiments on the Political Blogs dataset}\label{appx:exp_poliblog}

While the original graph has 1490 nodes, we followed standard practice in selecting the largest (weakly) connected component of the graph, which contains 1222 nodes. We denote this graph as $G$.The true partition of the blogs according to political leaning is available, denoted $x_{\mathrm{True}}$ here. This also allows accurate estimates of the graph parameters $(a,b)$ to be made, and we use these estimates for $a,b$ for GoF, and for the semi-synthetic procedure for TST. We found that $\hat a \approx 49.5$, while $\hat b \approx 5.2$, giving a ratio $a/b \approx 10$. The communities, according to $x_{\mathrm{True}}$ are of sizes 636 and 586.

The regime of low $\Lambda$ is explored via sparsification. Fixing a $\rho \in (0, 1]$, sparsification is performed by independently flipping coins for each edge in $G$, and keeping the edge with probability $\rho$. We refer to $\rho$ as the rate of sparsification.

We lastly note that at no sparsification ($\rho = 1$), spectral clustering produces a partition $\hat{x}_1$ such that $d(x_{\mathrm{True}}, \hat{x}_1) = 56$. 

\paragraph{GoF Procedure.} \begin{enumerate}
    \item The graph is sparsified at rate $\rho$. Let the sparsened graph be $G_\rho$.
    \item For the na\"{i}ve recovery based scheme, spectral clustering is performed on $G_\rho$ as in the previous section to generate $\hat{x}_\rho$.
    \item For the proposed test from \S\ref{sec:gof}, the statistic is computed on $G_\rho$. 
    \item The size of the test is estimated by running the GoF tests with $x_0 = x_{\mathrm{True}}.$ For the na\"{i}ve scheme, we reject if $d(\hat{x}_\rho, x_0) \ge s / 2$; for the proposed scheme, we use the test from \S \ref{sec:gof}. 
    \item To compute the power at distortion $s$, we first generate $y$ by randomly inverting the community labels of $s$ nodes in $x_{\mathrm{True}}$. We then run the same procedure as in the previous line, but with $x_0 = y$. Note that the graphs are not edited in any way.
    \item The precise implementation details are exactly as in Appendix \ref{appx:sbm-expt-details}, with the minor difference that we use a regularizer of $\tau = 1$ for spectral clustering.
\end{enumerate}

\paragraph{TST Proceudre.}

\begin{enumerate}
    \item Recall that TST requires two graphs as input. The experiment compares the political blogs graph against SBMs.
    \item To compute the size, we require a graph with the same underlying communities as $G$. Thus we generate $G',$ which is drawn as an SBM with the underlying partition $x_{\mathrm{True}}$, and parameters $a,b$ as estimated from the political blogs graph $G$.
    \item To determine the power of the tests we need a graph with an $s$-far underlying community. For this, we first generate a $y$ such that $d(x_{\mathrm{True}}, y) = s$, as we did in the GoF Procedure. Next, we sample $H$ as an SBM with underlying partition $y.$
    \item The graphs $G$, $G'$ and $H$ are now all sparsified at rate $\rho$ to get $G_\rho$, $G'_\rho$ and $H_\rho$.
    \item The size of each test is estimated using the TST procedures, as described in Appendix \ref{appx:sbm-expt-details} on the pair $(G_\rho, G'_\rho)$. Power is similarly estimated using the TST procedures on the pair $(G_\rho, H_\rho)$.
\end{enumerate}

\subsection{Experiments on the GMRFs}\label{appx:exp_gmrf}

Following the heuristic detailed in \S\ref{sec:exp_gmrf}, we na\"{i}vely generalise community recovery and testing to this setting, by replacing all instances of the graph adjacency matrix in previous settings with the sample covariance matrix. 

\label{appx:gmrf-expt-details}

The Gaussian Markov Random Field is described by its precision matrix $\Theta$ (i.e., the inverse covariance matrix of the Gaussian random vector on its nodes). We perform a preliminary examination of the possibility of testing changes in communities for an SBM-structured GMRF even when learning the structure is hard or impossible. As described in Section~\ref{sec:exp_gmrf}, we set
\begin{equation*}
    \Theta = I + \gamma G,
\end{equation*}
where G is the adjacency matrix of an SBM with known parameters. We generate samples from the GMRF as follows:
\begin{enumerate}
    \item For a fixed number of nodes $n$, we fix an $\SNR$ for the SBM, $\Lambda,$ and compute $(a, b)$ satisfying this $\Lambda$ so that $b / a = r$.
    \item Here, we consider $n=1000$ nodes and take $\Lambda = 30 \Lambda_0$, where $\Lambda_0 = \frac{10}{11} \log(n / 100) \approx 2.1$, as before, and $r = 1/10$. We find that $(a, b) \approx (12.34 \log n, 1.234 \log n)$. Note that since $\Lambda \approx 63 \approx 10\log(n),$ recovery of the communities for a raw SBM at this SNR is trivial.
    \item We fix a GMRF parameter $\gamma$. Here, we take $\gamma = 3 / (a + b) \approx 0.032$.
    \item We can now construct the precision matrix $\Theta$ after sampling $G$ from the SBM. We re-sample to ensure that $\Theta$ is positive-definite, but in practice, for the value of $\gamma$ quoted above, we did not encounter the need to re-sample.
    \item To generate i.i.d.\ samples $\zeta \sim \mathcal N(0, \Theta^{-1})$ in a stable manner, we use the following algorithm:
    \begin{enumerate}
        \item Compute the lower-triangular Cholesky factor $R$ of $\Theta$, so that $\Theta = R R^\tpose$.
        \item Sample $\xi \sim \mathcal N(0, I)$ from a standard $n$-dimensional multivariate normal distribution.
        \item Solve for $\zeta$ in $R^\tpose \zeta = \xi$.
    \end{enumerate}
    This suffices, since, $\zeta = (R^\tpose)^{-1} \xi$ would then have the covariance matrix $(R^\tpose)^{-1} R^{-1} = (R R^\tpose)^{-1} = \Theta^{-1}$.
    \item In this manner, we generate samples from the null and alternate distributions: let $\zeta$, $\zeta'$ and $\upsilon$ respectively denote samples drawn from a GMRF structured using $G$, $G'$ and $H$ respectively. Here, $G$, $G'$ and $H$ exactly are as described in Section~\ref{appx:sbm-expt-details}.
\end{enumerate}

Next, we describe how each of the two schemes is evaluated:
\begin{enumerate}
    \item Assuming we have $t$ i.i.d.\ samples of $\zeta$, generated as described above, we estimate the covariance matrix $\hat \Sigma$ of $\zeta$ using the standard estimator:
    \begin{equation*}
        \hat \Sigma = \frac{1}{t - 1} \sum_{i=1}^t (\zeta_i - \bar\zeta) (\zeta_i - \bar\zeta)^\tpose,
    \end{equation*}
    where $\bar\zeta = \frac{1}{t}\sum_{i=1}^t \zeta_i$. We then compute the correlation matrix,
    \begin{equation*}
        \hat C : \hat C_{ij} = \frac{\hat\Sigma_{ij}}{\sqrt{\hat\Sigma_{ii} \hat\Sigma_{jj}}},
    \end{equation*}
    which will be used in place of the adjacency matrix for both two-sample testing schemes.
    \item Similarly, we compute $\hat C$, $\hat C'$ and $\hat D$ from $\zeta$, $\zeta'$ and $\upsilon$ respectively.
    \item The na\"ive two-sample test based on recovery and comparison is evaluated exactly as described in Section~\ref{appx:sbm-expt-details}, except that $\hat C$, $\hat C'$ and $\hat D$ are used in place of $G$, $G'$ and $H$ respectively. False alarm and missed detection rates are also computed in exactly the same way.
    \item The two-sample test based on Algorithm~\ref{alg:tst} has several important variations:
    \begin{enumerate}
        \item We use the test statistics
        \begin{align*}
            T_{\text{Null}} &= T^{\hat x}(\hat C) - T^{\hat x}(\hat C') \\
            T_{\text{Alt.}} &= T^{\hat x}(\hat C) - T^{\hat x}(\hat D),
        \end{align*}
        for the null and alternate distributions respectively. Here, $\hat x$ has been estimated from $\hat C$.
        \item The threshold for the test is estimated from data. That is, we simulate $M=100$ samples of $T_{\text{Null}}$ and $T_{\text{Alt.}}$ each, and fit a classifier to differentiate between the two distributions. The classifier used is a simplistic 1-dimensional Linear Discriminant Analysis.
        \item We estimate false alarm and missed detection rates by applying the classifier to a hold-out dataset. To use the data as efficiently as possible, we use 10-fold repeated, stratified cross-validation, with 10 repetitions.
    \end{enumerate}
\end{enumerate}

\paragraph{Remark on subsampling.}
\begin{enumerate}
    \item Note that in the two-sample test for GMRFs based on Algorithm~\ref{alg:tst}, we do not subsample $\hat C$ as we did before in the case of SBMs.
    \item While previously, we had subsampled $G$ to create two subgraphs $G_1$ and $\tilde G$ that shared independence properties for ease of theoretical analysis, it should be noted that subsampling results in an effective loss of SNR. This is also the reason why we had to adjust the implementation using a different rate $\eta$.
    \item However, it emerges empirically that skipping the subsampling entirely, with a completely dependent $\hat x$ and $G$, makes for better separation between the null and alternate distributions, providing a more powerful statistic.
    \item Since we could not analytically derive a threshold for this statistic, we presented the subsampled test statistic for the first experiment.
    \item Since in the case of GMRFs, we are estimating the threshold from data, we use the more powerful test statistic to show the full extent of possible gains when using a dedicated algorithm for change detection, instead of na\"ively looking for changes by learning community structures first. 
\end{enumerate}

\end{appendix}
\end{document}